\documentclass{aa}  

\usepackage{graphicx}
\usepackage{xcolor}
\usepackage{subfigure}          
\usepackage{txfonts}
\usepackage{hyperref}           
\usepackage{arydshln}           

\usepackage{comment}            
\usepackage{acronym}            

\acrodef{ALMA}[ALMA]{Atacama Large Millimeter/Submillimeter Array}
\acrodef{JWST}[JWST]{James Webb Space Telescope}
\acrodef{MIRI}[MIRI]{Mid-Infrared Instrument}
\acrodef{MINDS}[MINDS]{MIRI Mid-Infrared Disk Survey}
\acrodef{XEST}[XEST]{XMM-Newton Extended Survey of the Taurus Molecular Cloud}
\acrodef{MHD}[MHD]{magnetohydrodynamic}
\acrodef{pim}[pim]{pebble isolation mass}

\begin{document}

\title{Changing disc compositions via internal photoevaporation II: \\ M dwarf systems}

\author{J. L. Lienert\inst{1}\fnmsep\thanks{Corresponding author; \email{lienert@mpia.de}}, B. Bitsch\inst{2} \and Th. Henning\inst{1}}

\institute{Max-Planck-Institut für Astronomie, Königstuhl 17, D-69117 Heidelberg, Germany \and University College Cork, College Rd, Cork T12 K8AF, Ireland}

\date{Received January 26, 2025; accepted May 27, 2025}

\abstract{The chemical evolution of the inner regions of protoplanetary discs is a complex process. It is influenced by several factors, one being the inward drift and evaporation of volatile-rich pebbles that can enrich the inner disc with vapour. During the evolution of the disc, its inner part is first enriched with evaporating water-ice, resulting in a low C/O ratio. Afterwards, carbon-rich gas from the outer disc, originating from the evaporation of CO, CO$_2$ and CH$_4$ ice, is viscously transported inwards, while at the same time, the supply of water-rich pebbles ceases and the water vapour in the inner disc is accreted onto the star. Consequently, the C/O ratio of the inner disc increases again after $2 \, \text{Myr}$. Previously, we studied how internal photoevaporation influences the chemical composition and evolution of discs around Sun-like stars by carrying away gas and opening gaps that block inward drifting pebbles. We now extend our study to lower-mass stars ($M_{\star} = 0.1 - 0.5 \, \text{M}_{\odot}$), where the time evolution of the disc's C/O ratio is different due to the closer-in position of the evaporation fronts and differences in disc mass, size and structure. Our simulations are carried out with a semi-analytical 1D disc model. The code \texttt{chemcomp} includes viscous evolution and heating, pebble growth and drift, pebble evaporation and condensation, as well as a simple chemical partitioning model for the disc. We show that internal photoevaporation plays a major role in the evolution of protoplanetary discs and their chemical composition: As for solar-mass stars, photoevaporation opens a gap, which stops inward drifting pebbles. They then cannot contribute to the volatile content of the gas in the inner disc any more. In addition, volatile-rich gas from the outer disc, originating from evaporated CO, CO$_2$ or CH$_4$ ice, is carried away by the photoevaporative winds. Consequently, the C/O ratio in the inner disc remains low, contradicting observations of the composition of discs around low-mass stars. Our model implies that inner discs at young ages ($< 2 \, \text{Myr}$) should be oxygen-rich and carbon-poor, while older discs ($> 2 \, \text{Myr}$) should be carbon-rich. The survival of discs to this age can be attributed to lower photoevaporation rates. These lower rates could either originate from the large spread of observed X-ray luminosities or from the photoevaporation model used in this study \citep[based on][]{picognaDispersalProtoplanetaryDiscs2021b}, which likely overestimates the photoevaporation efficiency at a given X-ray luminosity, leading to discrepancies with the observed C/O ratios in discs around low-mass stars. A reduction of the photoevaporation rate brings the calculated elemental abundances into better agreement with observations.}


\maketitle


\section{Introduction} \label{sec:Introduction}
In recent years, a lot of observational progress has been made in detecting and characterising exoplanets. It is therefore well known that planetary systems show a large diversity in their architectures and planetary properties. However, the exact pathways to end up in the observed final states are still investigated \citep[for a review, see][]{drazkowskaPlanetFormationTheory2022}. Since only very few observations of systems with forming planets exist \citep{kepplerDiscoveryPlanetarymassCompanion2018,mullerOrbitalAtmosphericCharacterization2018}, it is crucial to study their formation environment via simulations. That way, we can gain greater insight into the structure and composition of protoplanetary discs, helping us to constrain the properties and formation pathways of planets. \\
Planetary cores are believed to form via solid accretion in the so-called core accretion scenario. The solids can either be accreted in the form of planetesimals \citep{idaDeterministicModelPlanetary2004,guileraPlanetesimalFragmentationGiant2014,mordasiniPlanetaryPopulationSynthesis2018,miguelDiverseOutcomesPlanet2019,emsenhuberNewGenerationPlanetary2021} or pebbles \citep{ormelEffectGasDrag2010,lambrechtsRapidGrowthGasgiant2012,bitschGrowthPlanetsPebble2015,bitschFormationPlanetarySystems2019,nduguPlanetPopulationSynthesis2018,lambrechtsFormationPlanetarySystems2019,liuSuperEarthMassesSculpted2019,izidoroFormationPlanetarySystems2021}, with pebble accretion being more efficient than planetesimal accretion in the outer disc regions \citep[e.g.][]{johansenExploringConditionsForming2019}. Many population synthesis studies have used the core accretion scenario to try to reproduce observed masses and orbital distances of exoplanets. To constrain planet formation models even further, we can now use newly measured atmospheric abundances as well \citep[e.g.][]{molliereRetrievingScatteringClouds2020,lineSolarSubsolarMetallicity2021,pelletierWhereWaterJupiterlike2021,augustConfirmationSubsolarMetallicity2023}. This has been done in recent theoretical studies from \cite{mordasiniIMPRINTEXOPLANETFORMATION2016}, \cite{boothChemicalEnrichmentGiant2017a}, \cite{schneiderHowDriftingEvaporating2021,schneiderHowDriftingEvaporating2021a}, \cite{bitschHowDriftingEvaporating2022}, \cite{molliereInterpretingAtmosphericComposition2022} or \cite{penzlinBOWIEALIGNHowFormation2024}. \\
The final composition of a planet depends crucially on the composition of the disc. Therefore, simulations of protoplanetary discs need to include the full evolution of the disc's structure and chemical composition. A lot of factors play a role in determining the latter, such as the disc temperature and density, radiation fields, and the position of evaporation lines of different species \citep[see e.g.][]{cuzziMaterialEnhancementProtoplanetary2004,obergEFFECTSSNOWLINESPLANETARY2011,henningChemistryProtoplanetaryDisks2013,schneiderHowDriftingEvaporating2021,eistrupChemicalEvolutionIces2022,molliereInterpretingAtmosphericComposition2022}. The structure of the disc is in turn influenced by pressure bumps that can block inward drifting pebbles \citep[e.g.][]{pinillaTrappingDustParticles2012}, where the latter then plays a role again in the time evolution of the disc's composition. Indeed, \ac{JWST} observations by \cite{banzattiJWSTRevealsExcess2023}, \cite{grantMINDSDetection132023}, \cite{taboneRichHydrocarbonChemistry2023} and \cite{gasmanMINDSInfluenceOuter2025} have shown that pressure bumps influence the chemical composition of inner protoplanetary discs. This effect is seen in simulations as well, see works by \cite{bitschDryWaterWorld2021}, \cite{kalyaanEffectDustEvolution2023} or \cite{mahMindGapDistinguishing2024}. \\
One possible mechanism for causing these pressure bumps is a planet reaching its pebble isolation mass and creating a gap in the protoplanetary disc \citep[e.g.][]{paardekooperDustFlowGas2006,lambrechtsSeparatingGasgiantIcegiant2014,ataieeHowMuchDoes2018,bitschPebbleisolationMassScaling2018}. Two other mechanisms that significantly influence the composition and evolution of protoplanetary discs are \ac{MHD} winds \citep[for a review, see][]{lesurHydroMagnetohydroDustGas2023} or photoevaporation \citep[for a review, see][]{pascucciRoleDiskWinds2022a}. Both of these processes carry away disc material, which, in the case of photoevaporation, carves deep gaps into the disc. Since photoevaporation determines the final stages of disc evolution \citep[e.g.][]{ercolanoXRAYIRRADIATEDPROTOPLANETARY2009,pascucciEVIDENCEDISKPHOTOEVAPORATION2009,ercolanoMetallicityPlanetFormation2010,owenTheoryDiscPhotoevaporation2012a,owenCharacterizingThermalSweeping2013}, it is especially important as we know from observations that discs on average only live for a few million years \citep{mamajekInitialConditionsPlanet2009,fedeleTimescaleMassAccretion2010} and therefore need a disc dissolving mechanism. The specific lifetime of discs, however, can vary significantly depending on the properties of the central star of the system \citep[e.g.][]{michelBridgingGapProtoplanetary2021,pfalznerMostPlanetsMight2022}. \\
The process of photoevaporation describes the transfer of energy from high-energy photons to gas particles in the disc such that the latter increase their velocities so much that they escape the system. Depending on the source of the energetic radiation, we either refer to the evaporation process as external photoevaporation \citep[for a review, see][]{winterExternalPhotoevaporationPlanetforming2022}, when the radiation comes from nearby stars outside the studied system, or internal photoevaporation \citep[for a review, see][]{pascucciRoleDiskWinds2022a}, when the radiation's origin is the host star of the studied system. After enough gas has escaped the protoplanetary disc, a gap opens, resulting in a pressure bump in the disc. The gap blocks inward-moving pebbles and, at the same time, cuts off the inner regions from their gas supply from the outer disc as the photoevaporative winds continuously carry the gas away. \\
This work is a follow-up on our first paper, see \cite{lienertChangingDiscCompositions2024a} (hereafter referred to as Paper I), where we combined a viscous disc evolution model, including pebble drift and evaporation, with a model of internal photoevaporation and studied protoplanetary discs around solar-mass stars. Here, we extend this study to discs around lower-mass stars ($M_{\star} = 0.1 - 0.5 \, \text{M}_{\odot}$), especially focusing on the influence of the photoevaporative mass loss rate on the evolution of the elemental composition of the inner disc, which happens on faster timescales compared to Sun-like stars \citep[e.g.][]{mahCloseiceLinesSuperstellar2023}. Here and for the rest of the paper, the "inner disc" is defined as the part of the disc close to the star, separated from the disc's outer parts by the photoevaporative gap. The gap's inner edge, which defines the outer boundary of the inner disc, is generally located between $0.7 - 3 \, \text{AU}$, depending on the stellar mass and the photoevaporation rate. \\
This paper is structured as follows: Section \ref{sec:Methods} includes a summary of our model. The obtained results are shown in section \ref{sec:Results}, with their implications being discussed in section \ref{sec:Discussion}, before we conclude with section \ref{sec:Summary_and_conclusions}.

\section{Methods} \label{sec:Methods}
\begin{figure}
    \centering
    \begin{minipage}{0.5\textwidth}
        \includegraphics[width=\textwidth]{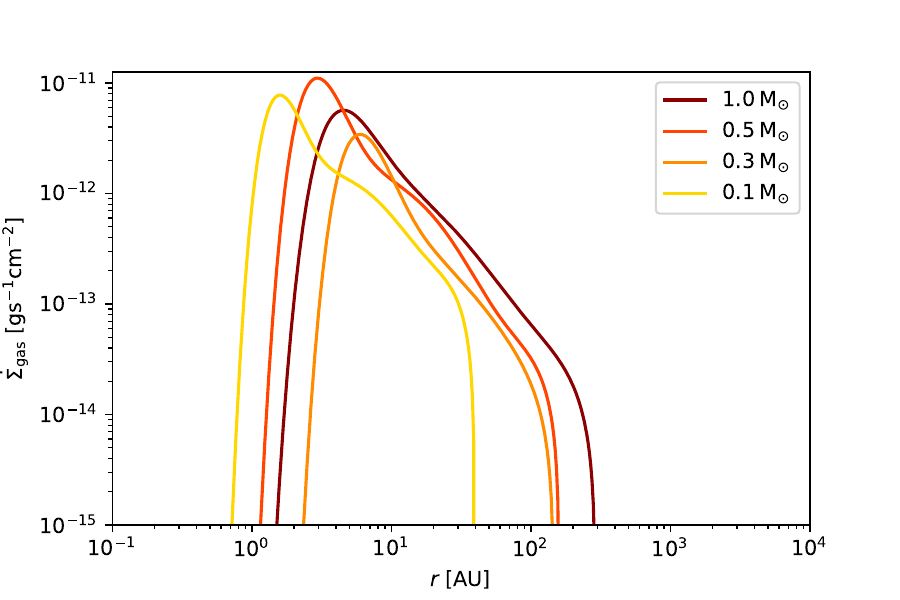}
    \end{minipage}
    \caption{Photoevaporative gas surface density loss rate as a function of disc radius for different stellar masses, as given in equation \ref{eq:sig_dot_ph}, adopted from \cite{picognaDispersalProtoplanetaryDiscs2021b}. Here, the nominal values for the photoevaporative mass loss rate, as given in table \ref{table:fit_parameters}, are used.}
    \label{fig:sig_dot}
\end{figure}

\begin{figure}
    \centering
    \begin{minipage}{0.5\textwidth}
        \includegraphics[width=\textwidth]{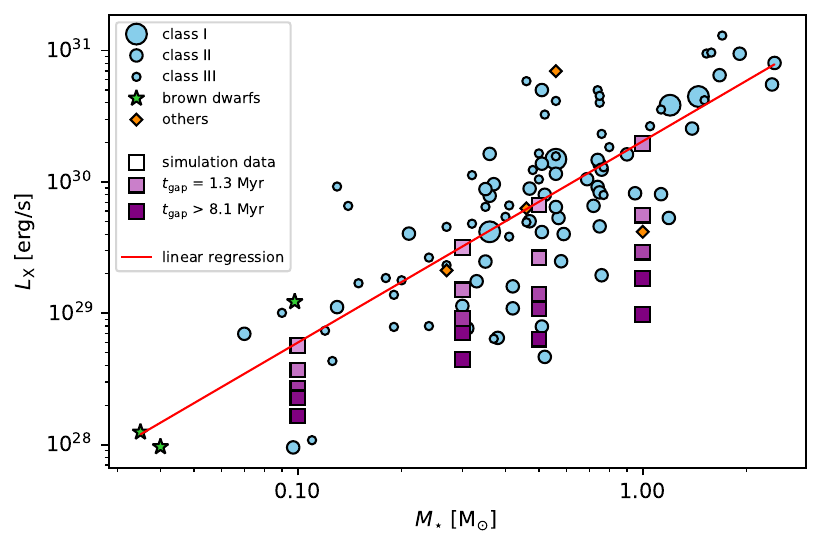}
    \end{minipage}
    \caption{X-ray luminosity, derived from the measured X-ray flux, as a function of stellar mass for all detected \ac{XEST} sources, re-plotted from \cite{gudelXMMNewtonExtendedSurvey2007} using their original data. The distance to Taurus is estimated with $140 \, \text{pc}$, although we note that Taurus consists of subgroups with varying distances \citep{galliStructureKinematicsTaurus2019}. The flux of stars with more than one measurement has been averaged to retrieve only one data point for the plot. The circle, star and diamond symbols define the stellar class, as defined for \ac{XEST} sources, and the straight line gives a linear regression for the logarithmic values, being described by the following equation, $\log(L_{\text{X}}) = 1.54 \log(M_{\star}) + 30.31$. Synthetic values from our simulations from sections \ref{ssec:Reduced_photoevaporation_rate} and \ref{ssec:Disc_lifetimes} are added as purple coloured squares, with the colour indicating the gap opening time of the disc, $t_{\text{gap}}$.}
    \label{fig:guedel_fig_12}
\end{figure}

\begin{table*}
    \centering
    \scalebox{0.93}{
	\begin{tabular}{cccccccccc}
		\hline
		$M_{\star}$ $[\text{M}_{\odot}]$	& $a$	& $b$	& $c$	& $d$	& $e$	& $f$	& $g$	& $\dot{M}_{\text{w}}$ $[10^{-8} \, \text{M}_{\odot} \text{yr}^{-1}]$  & $L_{\text{X}}$ $[10^{29} \, \text{erg s}^{-1}]$    \\
            \hline \hline
            $1.0$	& $-0.6344$	& $\phantom{0}6.3587$ & $-26.1445$    & $\phantom{0}56.4477$  & $-\phantom{0}67.7403$ & $43.9212$ & $-13.2316$            & $3.86446$ & $20.42$           \\
            $0.5$	& $-1.2320$	& $10.8505$           & $-38.6939$    & $\phantom{0}71.2489$  & $-\phantom{0}71.4279$ & $37.8707$ & $-\phantom{0}9.3508$  & $1.90460$ & $\phantom{0}7.02$ \\
            $0.3$   & $-1.3206$ & $13.0475$           & $-53.6990$    & $117.6027$            & $-144.3769$           & $94.7854$ & $-26.7363$            & $1.17156$ & $\phantom{0}3.20$ \\
            $0.1$	& $-3.8337$	& $22.9100$           & $-55.1282$    & $\phantom{0}67.8919$  & $-\phantom{0}45.0138$ & $16.2977$ & $-\phantom{0}3.5426$  & $0.37588$ & $\phantom{0}0.59$ \\ \hline
	\end{tabular}
    }
    \caption{Fit parameters for the photoevaporative gas surface density profile, taken from \cite{picognaDispersalProtoplanetaryDiscs2021b}. The $\dot{M}_{\text{w}}$ here are referred to as "nominal values" and represent the $\dot{M}_{\text{w}} (M_{\star},L_{\text{X,soft}})$ from equation \ref{eq:M_dot_ph}. Additionally shown are the X-ray luminosities for the different stellar masses, calculated with equation \ref{eq:X-ray_lum}.}
    \label{table:fit_parameters}
\end{table*}

\begin{table}
    \centering
	\begin{tabular}{ccc}
		\hline
		$M_{\star}$ $[\text{M}_{\odot}]$	& Reduction factor for $L_{\text{X}}$  & $\dot{M}_{\text{w}}$ $[10^{-8} \, \text{M}_{\odot} \text{yr}^{-1}]$	\\
		\hline \hline
            $0.5$	& $5.0$	& $0.51324$	\\
            $0.3$	& $3.5$	& $0.32733$	\\
            $0.1$	& $2.2$	& $0.10368$	\\ \hline
	\end{tabular}
    \caption{Photoevaporative mass loss rates (right column) for our stellar sample, reduced by a factor of 3-4 compared to the nominal values in table \ref{table:fit_parameters}. The reduced photoevaporation rates result from the reduction of the X-ray luminosity, with the reduction factor given in the middle column.}
    \label{table:lower_PE_rates}
\end{table}

\begin{table}
    \centering
	\begin{tabular}{cc}
	\hline
        Simulation parameter                        & Value                                     \\ \hline \hline
        Stellar mass $M_{\star}$             & $[0.5,0.3,0.1] \, \text{M}_{\odot}$       \\
        Stellar luminosity $L_{\star}$       & $[0.34,0.17,0.03] \, \text{L}_{\odot}$    \\
        Viscous parameter $\alpha$                  & $10^{-4}$                                 \\
        Initial disc mass $M_{\text{disc}}$  & $[0.05,0.03,0.01] \, \text{M}_{\odot}$    \\        
        Initial disc radius $R_{\text{disc}}$& $[95,65,30] \, \text{AU}$                 \\
        Initial dust-to-gas ratio                   & $2 \%$                                    \\ \hline
	\end{tabular}
    \caption{List of parameters used for our standard simulations.}
    \label{table:simulation_parameters}
\end{table}

Our numerical simulations are carried out with the code \texttt{chemcomp}, as described in \cite{schneiderHowDriftingEvaporating2021}, with added internal photoevaporation using the model of \cite{picognaDispersalProtoplanetaryDiscs2021b}. \texttt{Chemcomp} is a 1D, semi-analytical model of protoplanetary discs combined with a planetary growth model. For more details on the planet formation model, the interested reader is referred to \cite{schneiderHowDriftingEvaporating2021}. In this paper, we mostly use the disc module of the code, where the disc evolution follows a classic viscous evolution model \citep[see e.g.,][]{lynden-bellEvolutionViscousDiscs1974a}, using the $\alpha$-viscosity description of \cite{shakuraBlackHolesBinary1973}. \\
Dust grains in the disc can grow into pebbles, with their evolution being modelled by the two-population approach developed by \cite{birnstielSimpleModelEvolution2012}. This approach has originally been calibrated for solar-mass stars (\citeauthor{birnstielSimpleModelEvolution2012} \citeyear{birnstielSimpleModelEvolution2012}, see also the discussion in \citeauthor{pfeilTriPoDTriPopulationSize2024} \citeyear{pfeilTriPoDTriPopulationSize2024}). Without the calibration, our model might harbour small uncertainties in the grain growth and drift routine for lower-mass stars. The drift timescale of particles is directly dependent on their size. The particle's size in turn depends on the fragmentation velocity and the disc's viscosity. Both these parameters have large parameter ranges, with the fragmentation velocity being constrained to $1-10 \, \text{m/s}$ by experiments \citep{gundlachSTICKINESSMICROMETERSIZEDWATERICE2014}, while the turbulence parameter $\alpha$, which directly influences the viscosity, has a parameter range of $\alpha = 10^{-4} - 10^{-2}$ in usual disc settings (see e.g., \citeauthor{dullemondDiskSubstructuresHigh2018} \citeyear{dullemondDiskSubstructuresHigh2018} for the observational side, or \citeauthor{flockGapsRingsNonaxisymmetric2015} \citeyear{flockGapsRingsNonaxisymmetric2015} for the theory perspective). Therefore, the uncertainty in the combination of those two parameters easily exceeds the uncertainty coming from the calibration to a different stellar mass in the two-population approach by \cite{birnstielSimpleModelEvolution2012}. \\
The growth of the dust grains in the disc is limited by drift, fragmentation and drift-induced fragmentation. The relative ratio between large and small grains changes depending on the growth-limiting mechanism. It is $f = 0.75$ for the fragmentation limit and $f = 0.97$ for the drift limit. The grains are then advected with the average velocity of the two components weighted by their relative fractions. Enrichment of the disc happens through evaporating volatiles from inward drifting pebbles at the particular ice lines of each species \citep{schneiderHowDriftingEvaporating2021}.

\subsection{Viscous evolution} \label{ssec:Viscous_evolution}
The disc's viscous evolution is described by the viscous disc equation, which is given as the time evolution of the gas surface density $\Sigma_{\text{gas}}$ and can be derived from the conservation of mass and angular momentum \citep{pringleAccretionDiscsAstrophysics1981,armitageAstrophysicsPlanetFormation2013},
\begin{equation} \label{eq:viscous_disc_equation}
    \frac{\partial \Sigma_{\text{gas,}Y}}{\partial t} - \frac{3}{r} \frac{\partial}{\partial r} \left[ \sqrt{r} \frac{\partial}{\partial r} \left( \sqrt{r} \nu \Sigma_{\text{gas,}Y} \right) \right] = \dot{\Sigma}_{Y}.
\end{equation}
$\nu$ is the disc's viscosity and $\dot{\Sigma}_{Y}$ denotes the source term of the molecular species $Y$. For a full list of molecules included in the code, we refer to \cite{schneiderHowDriftingEvaporating2021} and Paper I. The source term results from the evaporation and condensation of pebbles and is given by
\begin{equation}
    \dot{\Sigma}_{Y} = \left\{
        \begin{matrix}
            \dot{\Sigma}_{Y}^{\text{evap}} \quad r <    r_{\text{ice,}Y} \\
            \dot{\Sigma}_{Y}^{\text{cond}} \quad r \geq r_{\text{ice,}Y}.
        \end{matrix}
    \right.
\end{equation}
Here, $\dot{\Sigma}_{Y}^{\text{evap}}$ and $\dot{\Sigma}_{Y}^{\text{cond}}$ are the source terms for evaporation and condensation of species $Y$. They originate from the evaporation and condensation of volatiles of molecule $Y$ at the respective ice line $r_{\text{ice,}Y}$. \\
The viscosity $\nu$ in equation \ref{eq:viscous_disc_equation} is given by
\begin{equation}
    \nu = \alpha \frac{c_\text{s}^2}{\Omega_{\text{K}}},
\end{equation}
where $\alpha$ describes the strength of the turbulence, $c_\text{s}$ is the isothermal sound speed and $\Omega_{\text{K}} = \sqrt{\frac{GM_{\star}}{r^3}}$ is the Keplerian angular frequency. Here, $M_{\star}$ denotes the mass of the host star and $r$ the radial distance to it. The dimensionless turbulence parameter $\alpha$ is kept fixed in time and has a radially constant value of $\alpha = 10^{-4}$. The speed of sound can be linked to the mid-plane temperature of the disc, which does not evolve in time for simplicity and is thus calculated only at the initialisation of the simulation via an equilibrium between viscous and stellar heating with radiative cooling. \\
One has to keep in mind that in reality, low-mass stars ($M_{\star} = 0.1 - 0.5 \, \text{M}_{\odot}$) will experience a decrease in their luminosity over time, leading to lower irradiation temperatures \citep{baraffeNewEvolutionaryModels2015}. As a result, the evaporation lines for different molecules will move inwards. However, the innermost evaporation lines might not be affected as their position is determined by viscous heating, which evolves slowly if the viscosity is low ($\alpha = 10^{-4}$). In addition, the inward flux of pebbles happens on much faster timescales than the luminosity evolution of the host star \citep{brauerCoagulationFragmentationRadial2008}. This means that the pebbles have already reached the inner disc and evaporated their volatile content before the shift in the evaporation lines would affect said evaporation. Therefore, keeping the mid-plane temperature constant in time is a reasonable assumption for our simulations, where we use a small turbulence parameter of $\alpha = 10^{-4}$, which results in a slow viscous evolution.

\subsection{Internal photoevaporation} \label{ssec:Internal_photoevaporation}
For the simulations in this paper, the viscous disc is subject to internal photoevaporation. In this case, an additional term $\dot{\Sigma}_{\text{w}}$, describing the photoevaporative loss rate of the gas surface density, is subtracted from the right-hand-side of the viscous disc equation \ref{eq:viscous_disc_equation},
\begin{equation}
    \frac{\partial \Sigma_{\text{gas,}Y}}{\partial t} - \frac{3}{r} \frac{\partial}{\partial r} \left[ \sqrt{r} \frac{\partial}{\partial r} \left( \sqrt{r} \nu \Sigma_{\text{gas,}Y} \right) \right] = \dot{\Sigma}_{Y} - \dot{\Sigma}_{\text{w}}.
\end{equation}
This additional term acts on the total gas surface density, not separately on each included molecule. The description we employ for photoevaporation due to X-rays follows the work of \cite{picognaDispersalProtoplanetaryDiscs2019a,picognaDispersalProtoplanetaryDiscs2021b} and \cite{ercolanoDispersalProtoplanetaryDiscs2021b}. Only the soft X-ray regime is used as it has shown to be the most efficient part of the X-ray spectrum. \\
The photoevaporative gas surface density loss rate is given by
\begin{align} \label{eq:sig_dot_ph}
    \dot{\Sigma}_{\text{w}} &= \left( \frac{6 a \ln (r)^5}{r \ln(10)^6} + \frac{5 b \ln (r)^4}{r \ln(10)^5} + \frac{4 c \ln (r)^3}{r \ln(10)^4} + \frac{3 d \ln (r)^2}{r \ln(10)^3} \right. \\
    &\quad \enspace + \left. \frac{2 e \ln (r)}{r \ln(10)^2} + \frac{f}{r \ln(10)} \right) \frac{\ln(10)}{2 \pi r} \dot{M}_{\text{w}}(r) \ [\text{M}_{\odot} \, \text{AU}^{-2} \, \text{yr}^{-1}], \nonumber
\end{align}
with a mass loss rate of
\begin{align} \label{eq:M_dot_ph}
    \dot{M}_{\text{w}} (r) = \quad &10^{a \log_{10} (r)^6 + b \log_{10} (r)^5 + c \log_{10} (r)^4 + d \log_{10} (r)^3} \\
    \quad \cdot &10^{e \log_{10} (r)^2 + f \log_{10} (r) + g} \cdot \dot{M}_{\text{w}} (M_{\star},L_{\text{X,soft}}) \ [\text{M}_{\odot} \text{yr}^{-1}], \nonumber
\end{align}
where the fit parameters $a-g$ for our stellar sample are presented in table \ref{table:fit_parameters}. $L_{\text{X,soft}}$ denotes the soft part of the X-ray luminosity of the star. \\
Equation \ref{eq:sig_dot_ph} is visualised in figure \ref{fig:sig_dot}, with the photoevaporative gas surface density loss rates being plotted as a function of disc radius and stellar mass. The mass loss peaks between $1 \, \text{AU}$ and $10 \, \text{AU}$, depending on the mass of the central star, with the general trend of a shift in the peak location towards larger radii with increasing stellar mass. The disc mass is generally lost between $0.7 \, \text{AU}$ and $300 \, \text{AU}$, enabling the opening of a gap in this disc region. Similar to the trend of the peak location, the gap position moves to larger disc radii as a function of stellar mass. \\
An exception from both these trends is seen for the $0.3 \, \text{M}_{\odot}$ star. The reason is that the photoevaporation rates result from a combination of two changes, namely the stellar mass and the X-ray luminosity. A change in stellar mass results in a different gravitational potential, which is responsible for keeping the gas close to the host star. The X-ray luminosity on the other hand is responsible for the repulsion of the gas. Changing it affects the internal physics of the underlying photoevaporation model. A lower X-ray luminosity, as in the case of the $0.3 \, \text{M}_{\odot}$ star, is associated with a harder spectrum in the soft X-ray regime, which leads to a shift of the peak location towards larger radii \citep{picognaDispersalProtoplanetaryDiscs2021b}. Additionally, the X-ray luminosity affects the thermal structure of the disc. Here, lower X-ray luminosities lead to a larger flaring of the disc, which allows mass removal from a larger region \citep{ercolanoDispersalProtoplanetaryDiscs2021b}. Combining all these effects leads to the non-uniform dependence of the photoevaporative gas surface density loss rates on the stellar mass that we see in figure \ref{fig:sig_dot}, because a change in stellar mass is always accompanied by a change in X-ray luminosity. A linear trend for both the peak and gap position of the photoevaporation rates is observed when fixing the X-ray luminosity across the stellar sample, see figure 9 from \cite{picognaDispersalProtoplanetaryDiscs2021b}, where the mass loss rate as a function of distance is then just dependent on the star's gravitational potential. \\
As we see from figure \ref{fig:sig_dot}, the photoevaporation rates vary in radius but are constant in time, leading to the same amount of disc material being taken away in each time step of the simulations. The evolution of other disc areas unaffected by photoevaporation is then solely determined by the viscous transport of gas. \\
The photoevaporative mass loss rate $\dot{M}_{\text{w}}(M_{\star},L_{\text{X,soft}})$ in equation \ref{eq:M_dot_ph} depends on the stellar mass and X-ray luminosity. According to \cite{gudelXMMNewtonExtendedSurvey2007}, the X-ray luminosity $L_{\text{X}}$ of a star can be calculated from its mass by using the following relation,
\begin{equation} \label{eq:X-ray_lum}
    \log_{10} \left( L_{\text{X}} \right) = \left(1.54 \pm 0.12 \right) \log_{10} (M_{\star}) + \left( 30.31 \pm 0.06 \right) \ [\text{erg s}^{-1}],
\end{equation}
where the stellar mass has to be given in multiples of the solar mass $\text{M}_{\odot}$. Employing equation \ref{eq:X-ray_lum}, we obtain the X-ray luminosities for our stellar sample, as given in the right column of table \ref{table:fit_parameters} \citep[see also table 1 in][]{picognaDispersalProtoplanetaryDiscs2021b}. \\
Since only the soft part of the X-ray spectrum ($0.1 - 1 \,$keV) is important for photoevaporation, we use the following scaling relation to calculate the soft X-ray luminosity from the full X-ray luminosity,
\begin{equation} \label{eq:soft_X-ray_lum}
    \log_{10} \left( L_{\text{X,soft}} \right) = 0.95 \log_{10} \left( L_{\text{X}} \right) + 1.19 \ [\text{erg s}^{-1}].
\end{equation}
This relation is a result of interpolating the values of table 4 in \cite{ercolanoDispersalProtoplanetaryDiscs2021b}. \\
The photoevaporative mass loss rate $\dot{M}_{\text{w}}(M_{\star},L_{\text{X,soft}})$ in equation \ref{eq:M_dot_ph} consists of two parts. One is its stellar mass-dependent part and the other one is its X-ray-dependent part. To calculate the first, stellar mass-dependent part, we employ equation (5) from \cite{picognaDispersalProtoplanetaryDiscs2021b},
\begin{equation}
    \dot{M}(M_{\star}) = 3.93 \cdot 10^{-8} \, M_{\star} \ [\text{M}_{\odot}/\text{yr}].
\end{equation}
The second, X-ray-dependent part is taken from \cite{ercolanoDispersalProtoplanetaryDiscs2021b},
\begin{equation}
    \log_{10} \left( \dot{M}(L_{\text{X,soft}}) \right) = A \exp \left( \frac{ \left( \ln \left( \log_{10} \left( L_{\text{X,soft}} \right) \right) - B \right)^2}{C} \right) + D \ [\text{M}_{\odot}/\text{yr}].
\end{equation}
The parameters are given as $A = -1.947 \cdot 10^{17}$, $B = -1.572 \cdot 10^{-4}$, $C = -2.866 \cdot 10^{-1}$, $D = -6.694$. \\
The combined mass loss rate is then given by a scaling relation because the stellar mass dependence needs to be rescaled by the difference in the soft X-ray component adopted with respect to the mean X-ray luminosity for a given stellar mass,
\begin{equation} \label{eq:M_dot_ph_combined}
    \dot{M}(M_{\star},L_{\text{X,soft}}) = \dot{M}(M_{\star}) \frac{\dot{M}(L_{\text{X,soft}})}{\dot{M}(L_{\text{X,soft,mean}})} \ [\text{M}_{\odot}/\text{yr}].
\end{equation}
Here, $L_{\text{X,soft}}$ is the soft part of the observed X-ray luminosity of any given input star, with the observational value taken from figure \ref{fig:guedel_fig_12} and the soft part being calculated via equation \ref{eq:soft_X-ray_lum}. To get $L_{\text{X,soft,mean}}$, we use the mass of our input star and calculate its mean X-ray luminosity by employing equation \ref{eq:X-ray_lum}. The soft part of that mean X-ray luminosity is then again calculated using equation \ref{eq:soft_X-ray_lum}. \\
Since there is a large spread in X-ray luminosities for one specific stellar mass, the actual X-ray luminosity of a star can differ massively from the mean value calculated with equation \ref{eq:X-ray_lum}, see figure \ref{fig:guedel_fig_12}. This results in a spread in the photoevaporative mass loss rates as well, which might change the lifetimes of the protoplanetary discs. To investigate how and to what extent the disc lifetimes change, we reduce the X-ray luminosities by a factor of $[5.0,3.5,2.2]$ compared to the mean values for our stellar sample of $[0.5,0.3,0.1] \, \text{M}_{\odot}$, see table \ref{table:lower_PE_rates}. These reduced X-ray luminosities are well within the spread of the observed X-ray luminosities of \cite{gudelXMMNewtonExtendedSurvey2007}, see simulation data in figure \ref{fig:guedel_fig_12}. The decrease in stellar X-ray luminosity leads to a reduction of the photoevaporative mass loss rates by a factor of $3-4$ compared to the nominal values. A decrease in the strength of internal photoevaporation will prolong the lifetimes of the protoplanetary discs, which is more in agreement with observations. \\
The simulations by \cite{sellekPhotoevaporationProtoplanetaryDiscs2024} show lower photoevaporation rates compared to \cite{picognaDispersalProtoplanetaryDiscs2021b} due to the inclusion of additional cooling processes. However, these studies have so far only been applied to solar-type stars. We thus probe reduced photoevaporation rates in discs around lower-mass stars by changing the X-ray luminosity, as explained above. \\
An additional factor not considered here is the time variability of X-ray luminosities. A star's luminosity is not constant over its lifetime, leading to a time-dependent photoevaporative mass loss rate. Furthermore, assuming a different disc structure, i.e. a puffed-up inner part, results in the extinction of X-ray radiation, hindering internal photoevaporation. The puff-up is a result of viscous heating, which is accounted for in our model \texttt{chemcomp}, but not in the photoevaporation models by \cite{picognaDispersalProtoplanetaryDiscs2019a}, they only incorporate stellar heating. Therefore, we cannot study how much photoevaporation would be hindered by puffed-up discs.

\subsection{Pebble evaporation and chemistry model} \label{ssec:Pebble_evaporation_and_chemistry_model}
Protoplanetary discs in our model consist of gas and dust grains. The grains can grow to pebbles with sizes ranging from millimetres to centimetres. Drift, turbulent fragmentation and drift-induced fragmentation limit their growth \citep{brauerCoagulationFragmentationRadial2008}. Both gas and dust move inwards through the disc, with the dust pebbles drifting faster due to gas drag in smooth discs without substructures \citep{mahMindGapDistinguishing2024}. In discs containing pressure bumps, pebbles move towards pressure maxima. On their inward journey through the disc, they cross several ice lines of different molecular species and evaporate their respective volatile content. Consequently, the inner disc is enriched with the evaporated material. When gas gets closer to the central star, it will eventually be accreted onto it. On the other hand, re-condensation at evaporation fronts is also possible for gaseous species when they are moving outwards rather than inwards \citep{rosIceCondensationPlanet2013}. This can lead to large enhancements of solids compared to the solar composition \citep[e.g.][]{aguichineMassRadiusRelationships2021,mousisSituExplorationAtmospheres2022,mahFormingSuperMercuriesRole2023}. Inward drifting pebbles are self-consistently stopped at pressure bumps. \\
For the initial composition of the disc, we assume solar abundances. The exact values can be found in Paper I and are taken from \cite{asplundChemicalCompositionSun2009}. The elements are then distributed into volatile and refractory molecules, with CO, CO$_2$ and CH$_4$ as the main carbon-bearing species and water as the main oxygen-bearer. We note that C$_2$H$_2$ may be an important molecular carrier of carbon in the gas phase as well (Grant et al. 2025, submitted), but is currently not included in our model. The distribution of elements $X$ into molecules $Y$ follows a simple chemical partitioning model from \cite{schneiderHowDriftingEvaporating2021}, which is based on \cite{bitschInfluenceSubSupersolar2020}. The disc composition is dependent on the disc radius since the disc temperature varies with the orbital distance from the central star. Depending on their condensation temperature and position in the disc, molecules from a certain species $Y$ are then either present in gaseous or solid form. The orbital distance where the disc's mid-plane temperature is equal to the evaporation/condensation temperature of a certain species is called the evaporation/ice line of that species. \\
The number ratio of two elemental species $X_1$ and $X_2$ is defined via
\begin{equation}
    X_1 / X_2 = \frac{m_{X_1}}{m_{X_2}} \frac{\mu_{X_2}}{\mu_{X_1}},
\end{equation}
where $m_{X_1}$ and $m_{X_2}$ are the mass fractions of the two elements and $\mu_{X_1}$ and $\mu_{X_2}$ are their atomic masses. In this paper, this definition is used to calculate the C/O, C/H, O/H and N/H ratios.

\subsection{Initial conditions} \label{ssec:Initial_conditions}
Our standard simulations are carried out using host star masses of $M_{\star} = [0.5,0.3,0.1] \, \text{M}_{\odot}$. The corresponding stellar luminosities are retrieved from the stellar evolution models of \cite{baraffeNewEvolutionaryModels2015}, where we choose the luminosities at $2.5 \, \text{Myr}$ for this work, to be consistent with Paper I. The resulting values for the luminosities are $L_{\star} = [0.34,0.17,0.03] \, \text{L}_{\odot}$\footnote{\href{https://perso.ens-lyon.fr/isabelle.baraffe/BHAC15dir/BHAC15_tracks+structure}{https://perso.ens-lyon.fr/isabelle.baraffe/BHAC15dir/BHAC15\_tr-acks+structure}, retrieved on 22 March 2024}. The initial disc in our standard simulations is characterised by a turbulence parameter of $\alpha = 10^{-4}$, an initial disc mass of $M_{\text{disc}} = [0.05,0.03,0.01] \, \text{M}_{\odot}$, an initial disc radius of $R_{\text{disc}} = [95,65,30] \, \text{AU}$ and an initial dust-to-gas ratio of $2 \%$, as listed in table \ref{table:simulation_parameters}. The initial disc mass is fixed at $10 \%$ stellar mass under the assumption that our simulated systems start to evolve at an earlier stage when the disc is thought to be more massive than what is seen in observations. The values for the initial disc radius are obtained using the scaling relation of \cite{mahCloseiceLinesSuperstellar2023} based on \cite{andrewsScalingRelationsAssociated2018}, where $R_{\text{disc}} = 50 \, \text{AU} \, (M_{\star} / 0.1 \, \text{M}_{\odot})^{0.7}$. To match the parameters from our first paper, we scale down the ones calculated with the above-mentioned relation to arrive at the ones listed before. The initial disc radius is generally defined as the point in the disc where the exponential cut-off of the initial gas surface density sets in. The latter is described by a profile with an exponential decay in the outer disc regions, starting at $R_{\text{disc}}$, see appendix \ref{asec:Pure_viscous_disc} for the gas surface density evolution of a purely viscous disc. We run all our simulations for $10 \, \text{Myr}$ with a
time step of $10 \, \text{yr}$. The disc composition is retrieved every
$0.1 \, \text{Myr}$.

\section{Results} \label{sec:Results}
In this work, we study the influence of internal photoevaporation on the (chemical) evolution of discs around low-mass stars, i.e. stars with masses of $M_{\star} = 0.1 - 0.5 \, \text{M}_{\odot}$. The results are found in this section, with the main focus on the contrast between the nominal values for photoevaporative mass loss from \cite{picognaDispersalProtoplanetaryDiscs2021b} and reduced mass loss rates. The results for the nominal values are found in section \ref{ssec:Nominal_values_for_photoevaporative_mass_loss}, whereas the results for the reduced rates are presented in section \ref{ssec:Reduced_photoevaporation_rate}. Section \ref{sec:Results} closes with comparing the disc lifetimes for different photoevaporation rates in section \ref{ssec:Disc_lifetimes}. \\
For comparison, the appendices contain additional material on non-photoevaporative discs, see appendix \ref{asec:Pure_viscous_disc} for the case of a purely viscously evolving disc or appendix \ref{asec:Viscous_disc_with_a_planet} for a disc with a giant planet forming in it. Additionally, we provide more information on photoevaporative discs in appendix \ref{asec:Photoevaporative_disc}, where we study a reduction factor of $10$ on the photoevaporative mass loss as well as a photoevaporative disc with a factor $10$ increased viscosity.

\subsection{Nominal values for photoevaporative mass loss} \label{ssec:Nominal_values_for_photoevaporative_mass_loss}
\begin{figure*}
    \centering
    \includegraphics[width=\textwidth]{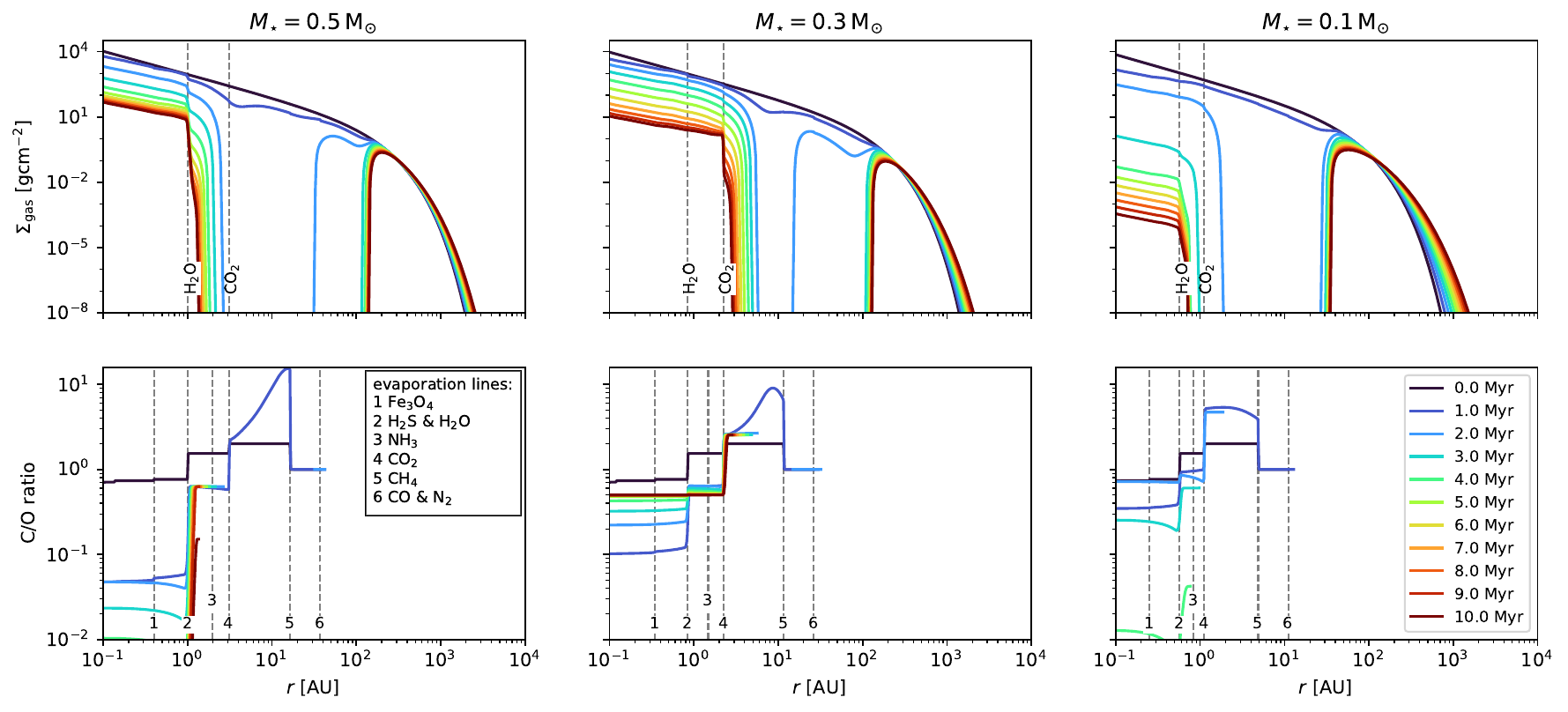}
    \caption{Disc evolution for a viscous disc with internal photoevaporation due to X-rays, using the nominal photoevaporative mass loss rates from table \ref{table:fit_parameters}. The host star masses vary from $0.5 \, \text{M}_{\odot}$ (on the left) to $0.1 \, \text{M}_{\odot}$ (on the right). \textbf{Top:} Gas surface density as a function of disc radius, time evolution is shown in colour - from black, which corresponds to $0 \, \text{Myr}$, to dark red, which corresponds to $10 \, \text{Myr}$. \textbf{Bottom:} Gaseous C/O ratio as a function of disc radius and time (colour-coded). The evaporation lines for different molecules are given as grey dashed lines. Note that the C/O ratio is calculated from number densities and that we, by definition, have no specified C/O ratio in the gas phase beyond the CO evaporation front. We use our standard parameters for this simulation, as given in table \ref{table:simulation_parameters}.}
    \label{fig:overview_disc_with_PE}
\end{figure*}

\begin{figure*}
    \centering
    \includegraphics[width=\textwidth]{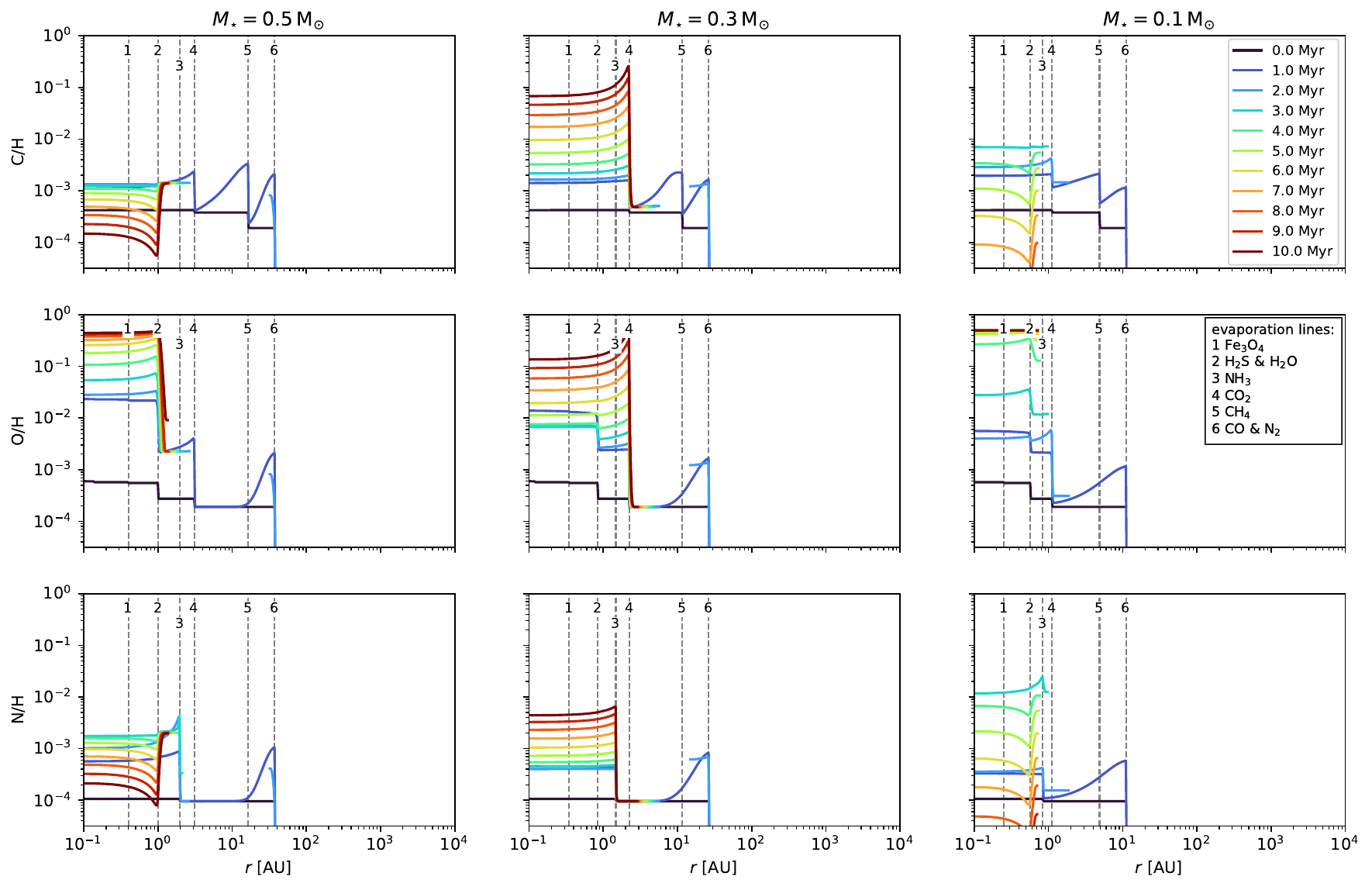}
    \caption{Different element ratios in the gas phase as a function of disc radius and time for a viscous disc with internal photoevaporation due to X-rays, using the nominal photoevaporative mass loss rates from table \ref{table:fit_parameters}. The host star masses vary from $0.5 \, \text{M}_{\odot}$ (on the left) to $0.1 \, \text{M}_{\odot}$ (on the right). \textbf{Top:} Carbon over hydrogen, \textbf{middle:} oxygen over hydrogen, \textbf{bottom:} nitrogen over hydrogen. Colour coding, plotting and simulation parameters as in figure \ref{fig:overview_disc_with_PE}.}
    \label{fig:element_ratios_disc_with_PE}
\end{figure*}

\subsubsection{Disc evolution}
All results shown in this subsection are obtained using the standard simulation parameters as described in section \ref{ssec:Initial_conditions} and shown in table \ref{table:simulation_parameters}. For the photoevaporative mass loss rates, we use the nominal values given in table \ref{table:fit_parameters}. \\
Figure \ref{fig:overview_disc_with_PE} shows the gas surface density in the top row and the gaseous C/O ratio in the bottom row, both as a function of disc radius, for a viscous disc with active internal photoevaporation. Stellar masses are varied across the different columns, with the mass decreasing from $M_{\star} = 0.5 \, \text{M}_{\odot}$ on the left to $M_{\star} = 0.1 \, \text{M}_{\odot}$ on the right. The time evolution spans over $10 \, \text{Myr}$, indicated via colour coding. The C/O ratio is calculated from number densities. \\
Our results for the gas surface density align with those from Paper I, where we conducted a similar study but for solar-mass stars. The gas surface density decreases over time in the inner disc due to viscous accretion of disc material onto the host star. In the outer disc regions, viscous spreading is the dominating mechanism. Both of these processes happen on relatively long timescales as the viscous parameter $\alpha$ is small, $\alpha = 10^{-4}$. \\
Additionally, internal photoevaporation opens a gap between $1-2 \, \text{Myr}$. Its inner edge moves inwards with time for all host star masses shown here. The exact gap size and position depend on the particular shape of the gas surface density loss rate, see figure \ref{fig:sig_dot}. Photoevaporation does not necessarily have the strongest visible effect at the position of the peak of the photoevaporative mass loss rate. Since the disc's surface density decays exponentially in its outer regions, it is important to check the relative strength of photoevaporation compared to the disc's gas surface density. \\
In comparison to paper I, the inner disc's gas surface density decreases faster for lower-mass stars, especially for the star with $M_{\star} = 0.1 \, \text{M}_{\odot}$. This is a direct result of the lower initial disc masses.

\subsubsection{Evolution of the C/O ratio}
The gap opened by photoevaporation has a large impact on the inner disc's chemical evolution. It blocks the pebbles from the outer disc regions, and at the same time, photoevaporative winds carry away inward-moving gas. The direct results of this mechanism are seen in the C/O ratio, see lower panels of figure \ref{fig:overview_disc_with_PE}. \\
The initial C/O ratio in figure \ref{fig:overview_disc_with_PE} shows the classical step-like behaviour \citep[e.g.][]{obergEFFECTSSNOWLINESPLANETARY2011,molliereInterpretingAtmosphericComposition2022}, with the different volatiles evaporating at different distances. These positions in the disc are determined by the disc temperature. The result is either an increase in the C/O ratio at the evaporation lines of carbon-rich molecules or a decrease in the C/O ratio at those of oxygen-rich molecules. \\
As the disc starts to evolve, a rapid drop in the C/O ratio in the inner disc, interior to the water-ice line, is seen during the first one million years for all host star masses shown here. The C/O ratio then continues to stay subsolar for the rest of the time evolution, in contrast to the cases of a purely viscously evolving disc or a disc hosting a giant planet, where the C/O ratio reaches supersolar values at the end of the time evolution \citep[see appendices \ref{asec:Pure_viscous_disc}, \ref{asec:Viscous_disc_with_a_planet} and the study by][]{mahCloseiceLinesSuperstellar2023}. However, apart from the C/O ratio generally being subsolar for all host star masses in the case of the photoevaporative discs that we are investigating here, the rest of the time evolution differs between the three different central star masses. \\
\underline{$M_{\star} = 0.5 \, \text{M}_{\odot}$:} In this case, the C/O ratio continues to decrease after the initial rapid drop. Its behaviour is a direct consequence of pebble drift and evaporation. The position of the water evaporation line, marked with number $2$ in figure \ref{fig:overview_disc_with_PE}, is at a smaller disc radius than those of the carbon-bearing molecules CO$_2$, CH$_4$ and CO (marked with $4$, $5$ and $6$). Consequently, pebbles from outer disc regions first reach the evaporation lines of the carbon-bearing species before crossing the water-ice line when moving inwards. Therefore, the carbon-rich vapour is created further out in the disc than the water vapour. Additionally, pebbles move much faster than gas, resulting in water-ice pebbles reaching the inner disc much earlier than the carbon-rich gas. As a result, the inner disc is first enriched with water vapour, leading to the rapid drop in the C/O ratio in the first one million years due to the large amount of oxygen contained in water. With time, the carbon-rich gas that formed in the outer disc regions enters the inner parts. At the same time, inner disc material is accreted onto the host star. Before a visible effect of these processes can occur, a gap opens in the disc between $1 \, \text{Myr}$ and $2 \, \text{Myr}$. The before-mentioned processes stop since neither pebbles nor gas can pass the deep gap carved by photoevaporation. Instead, another mechanism comes into play. \\
After the gap is fully opened, the flux of carbon vapour from the outer disc is blocked. On the other hand, the gas already present in the inner disc is accreted onto the host star, but with one exception. This is due to the occurrence of a water equilibrium cycle at the water evaporation front (see also Paper I for more details). Since the water-ice line lies at the outer edge of the inner disc (see gas surface density in the upper left panel of figure \ref{fig:overview_disc_with_PE}), water vapour diffuses outwards due to the strong pressure gradient caused by the photoevaporative gap. It can then recondense at the water evaporation front and form water-ice pebbles. These then drift inwards and evaporate again. This process enables water to stay in the inner disc on much longer timescales than carbon-bearing species that have their evaporation lines further out and not within the inner disc (for comparison see the position of the CO$_2$ ice line in the gas surface density plot in the upper left panel of figure \ref{fig:overview_disc_with_PE}). Carbon-bearing molecules instead are either accreted onto the central star or removed via outward diffusion into the photoevaporative gap. As a result, the C/O ratio decreases over time as oxygen remains very abundant in the inner disc but carbon continues to be removed. \\
\underline{$M_{\star} = 0.3 \, \text{M}_{\odot}$}: In this case, the C/O ratio increases after the initial rapid drop and continues to do so for the rest of the time evolution until it reaches a value of $0.5$. Whereas the explanation for the initial drop is the same as for the $0.5 \, \text{M}_{\odot}$ star, the following increase differs in its behaviour and explanation. Instead of the water equilibrium cycle, the $0.3 \, \text{M}_{\odot}$ star has a CO$_2$ equilibrium cycle since the CO$_2$ evaporation front instead of the water-ice line now lies at the outer edge of the inner disc (see upper middle panel of figure \ref{fig:overview_disc_with_PE}), where a strong pressure gradient leads to the outward diffusion of material. Due to its position closer to the photoevaporative gap, the CO$_2$ equilibrium cycle is dominating over the equilibrium cycle for water, which does not exist in this case. Therefore, the CO$_2$ is kept in the inner disc, leading to a continuous increase of the C/O ratio in this region. With time, all other molecules except CO$_2$ are accreted onto the host star, leading to a final value of $0.5$ for the C/O ratio. This value is a direct consequence of the distribution of atoms in the CO$_2$ molecule. \\
It should be noted that the CO$_2$ equilibrium cycle, although being essential in keeping the carbon in the inner disc over longer time periods, is very sensitive to the relative position of the inner edge of the photoevaporative gap and the evaporation lines. Changing for example the disc's viscosity, as we have done in appendix \ref{assec:Larger_viscous_parameter_alpha}, shifts the CO$_2$ evaporation line for the $M_{\star} = 0.3 \, \text{M}_{\odot}$ star into the gap, leading to the non-existence of the CO$_2$ equilibrium cycle. The gap position itself is in turn dependent on the photoevaporation model, as can be seen in figure \ref{fig:sig_dot}. Another important factor is the disc's thermal structure, which depends on viscous heating and the evolution of the stellar luminosity, and which also affects the positions of the evaporation lines. \\
\underline{$M_{\star} = 0.1 \, \text{M}_{\odot}$}: In this case, the C/O ratio increases after the initial drop up to super-solar values, where the initial drop is again a result of the inner disc first being enriched with water vapour. The following increase is a consequence of carbon-rich vapour being carried into the inner disc. This happens faster in comparison to the other stellar masses because the evaporation lines for the $0.1 \, \text{M}_{\odot}$ star are closer to the host star. Additionally, between $2 \, \text{Myr}$ and $3 \, \text{Myr}$, the existence of a CO$_2$ equilibrium cycle ensures that carbon-rich gas stays in the inner disc. At the same time, carbon-rich gas from the outer disc can no longer reach the inner disc because the gap is already fully opened. After $3 \, \text{Myr}$, the CO$_2$ equilibrium cycle ceases to exist. The reason is the inward movement of the inner edge of the gap, shifting the CO$_2$ ice line from being in the inner disc to now being inside the widened gap. In turn, the water-ice line now lays at the outer edge of the inner disc and a water equilibrium cycle occurs as in the case of the $0.5 \, \text{M}_{\odot}$ star. As a result, the C/O ratio decreases rapidly after $3 \, \text{Myr}$.

\subsubsection{C/H, O/H and N/H}
In addition to the C/O ratio, we similarly investigate the C/H, O/H and N/H abundances in the inner disc, see top, middle and bottom row of figure \ref{fig:element_ratios_disc_with_PE}. As for the C/O ratio, we show the time evolution of these element ratios as a function of disc radius for different stellar masses, indicated at the top of each column. \\
As expected, there is not much carbon and nitrogen in the case of the $0.5 \, \text{M}_{\odot}$ and $0.1 \, \text{M}_{\odot}$ stars but a lot of oxygen in their inner discs at the end of the time evolution. However, the $0.1 \, \text{M}_{\odot}$ star initially has a CO$_2$ and an NH$_3$ equilibrium cycle, resulting in an elevated C/H and N/H ratio after $3 \, \text{Myr}$ compared to the C/H and N/H ratios at the same time of the $0.5 \, \text{M}_{\odot}$ star. These equilibrium cycles however stop once the inner edge of the photoevaporative gap reaches the position of their ice lines, resulting in a decrease of the C/H and N/H ratio. For the $0.3 \, \text{M}_{\odot}$ star, the carbon and nitrogen fractions are generally much higher, which is in agreement with our results for the C/O ratio and the position of the CO$_2$ and the NH$_3$ ice line. In contrast to the $0.5 \, \text{M}_{\odot}$ and $0.1 \, \text{M}_{\odot}$ stars, these two ice lines in the case of the $0.3 \, \text{M}_{\odot}$ star keep their position in the inner disc throughout the full time evolution. Due to their position at the outer edge of the inner disc and the consequential existence of equilibrium cycles for both molecules, we end up with an elevated carbon and nitrogen content of the inner disc.

\subsection{Reduced photoevaporation rate} \label{ssec:Reduced_photoevaporation_rate}
\subsubsection{Gas surface density and C/O ratio}
\begin{figure*}
    \centering
    \includegraphics[width=\textwidth]{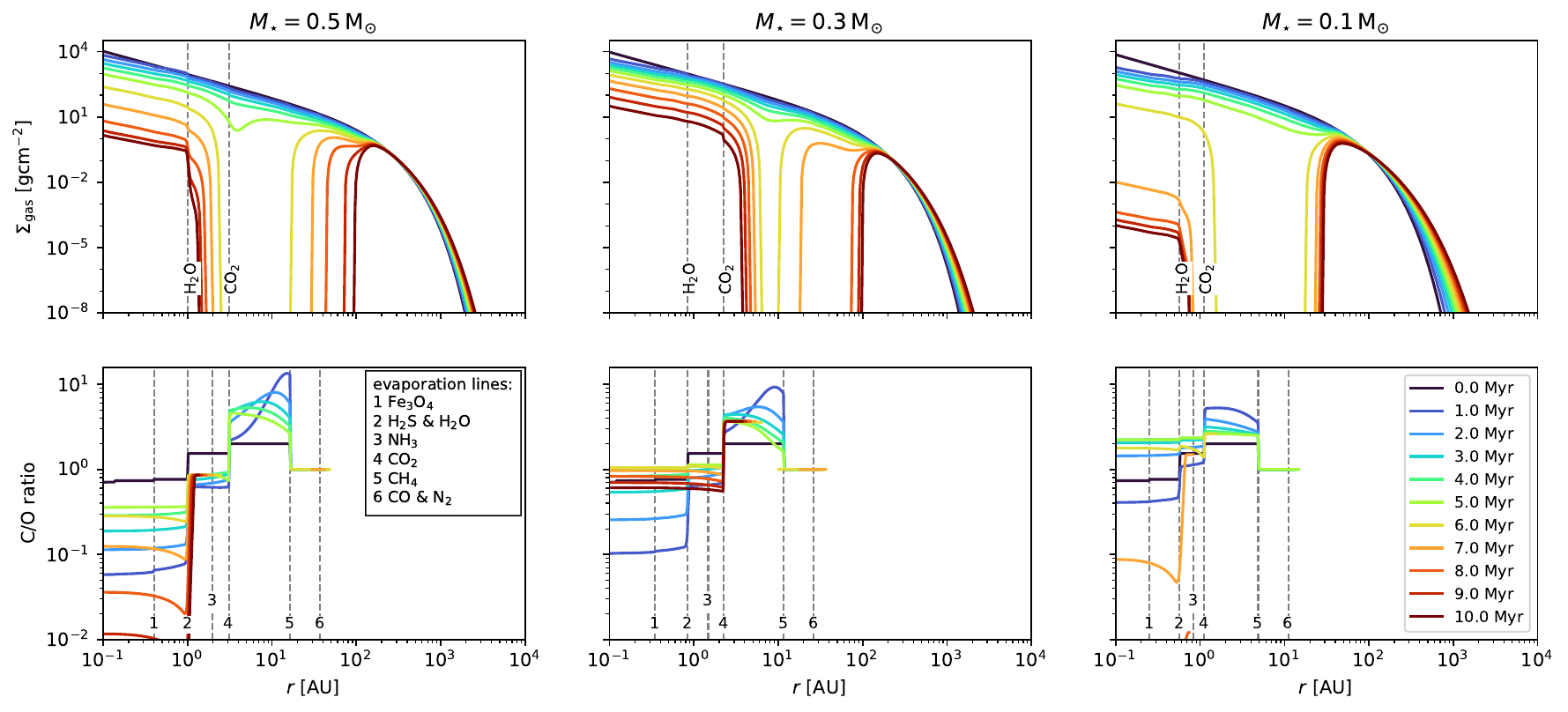}
    \caption{Disc evolution for a viscous disc with internal photoevaporation due to X-rays, but with factor 3-4 reduced mass loss rates. For the exact values, see table \ref{table:lower_PE_rates}. The reduced rates result in later gap opening and therefore longer disc lifetimes. Host star masses vary from $0.5 \, \text{M}_{\odot}$ (on the left) to $0.1 \, \text{M}_{\odot}$ (on the right). \textbf{Top:} Gas surface density as a function of disc radius and time. \textbf{Bottom:} Gaseous C/O ratio as a function of disc radius and time. Colour coding, plotting and simulation parameters as in figure \ref{fig:overview_disc_with_PE}.}
    \label{fig:overview_disc_with_factor_3to4_lower_PE}
\end{figure*}

\begin{figure*}
    \centering
    \includegraphics[width=0.8\textwidth]{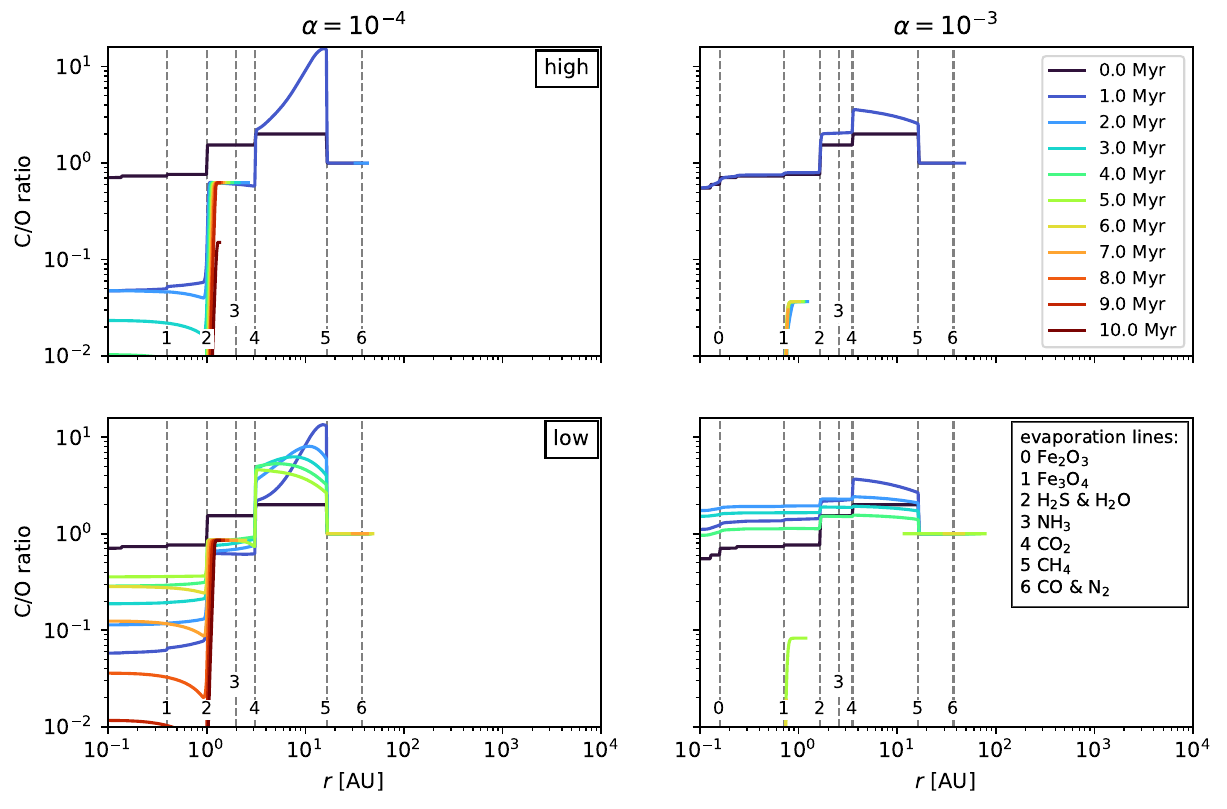}
    \caption{Gaseous C/O ratio as a function of disc radius and time for discs with internal photoevaporation due to X-rays and a host star mass of $0.5 \, \text{M}_{\odot}$. This plot shows a comparison between different $\alpha$ values and different photoevaporation rates, with the left column depicting $\alpha = 10^{-4}$, the right column depicting $\alpha = 10^{-3}$, the top rowing showing our nominal photoevaporation rate of $\dot{M}_{\text{w}} = 1.90460 \cdot 10^{-8} \, \text{M}_{\odot} \text{yr}^{-1}$ (see also table \ref{table:fit_parameters}) and the bottom row showing a reduced photoevaporation rate of $\dot{M}_{\text{w}} = 0.51324 \cdot 10^{-8} \, \text{M}_{\odot} \text{yr}^{-1}$ (see also table \ref{table:lower_PE_rates}). Colour coding, plotting and remaining simulation parameters as in figure \ref{fig:overview_disc_with_PE}.}
    \label{fig:C_to_O_ratio_comparison}
\end{figure*}

The results shown in this subsection are obtained using the standard simulation parameters as described in section \ref{ssec:Initial_conditions} and shown in table \ref{table:simulation_parameters}. For the photoevaporative mass loss rates, we do not use the nominal values, as in the previous subsection, but reduced rates as given in table \ref{table:lower_PE_rates}, with a reduction factor of $3-4$. A description of how these are obtained can be found in section \ref{ssec:Internal_photoevaporation}. \\
Figure \ref{fig:overview_disc_with_factor_3to4_lower_PE} shows the gas surface density in the top row and the gaseous C/O ratio in the bottom row, both as a function of disc radius, for discs with less effective internal photoevaporation. As in figure \ref{fig:overview_disc_with_PE}, time evolution is indicated via colour coding and the different host star masses are given at the top of each column. \\
The gas surface density generally behaves similarly to that shown in the previous subsection, where nominal values are used for the photoevaporative mass loss. We see a decrease in the inner disc due to accretion onto the central star and viscous spreading in the outer disc regions over time. However, the gap opened by photoevaporation occurs much later, at around $5-6 \, \text{Myr}$. Again, the $0.1 \, \text{M}_{\odot}$ star is the one with the highest mass loss in the inner disc. \\
The delayed gap opening directly impacts the C/O ratio, see lower panels of figure \ref{fig:overview_disc_with_factor_3to4_lower_PE}. As before, the initial phase is dominated by water-rich pebbles reaching the inner disc first, resulting in an excess of water vapour and therefore oxygen, which leads to a low C/O ratio. This phase is followed by carbon-rich vapour moving inwards, elevating the C/O ratio again. Since the gap opening is delayed in these simulations, the second phase is much longer than in the nominal case, resulting in much larger C/O ratios. The evolution of the C/O ratio during this phase follows that of a purely viscous disc, see figure \ref{fig:overview_viscous_disc}. After the gap has fully opened, the behaviour of the C/O ratio switches from following that of a purely viscous disc to that described in the previous subsection. In the case of the $0.5 \, \text{M}_{\odot}$ and $0.1 \, \text{M}_{\odot}$ stars, the inner disc is dominated by the water equilibrium cycle. As a result, their C/O ratios drop again quickly after gap opening. However, the drop of the C/O ratio is not as extreme as for the case of higher photoevaporation rates, compare the bottom row of figure \ref{fig:overview_disc_with_factor_3to4_lower_PE} for the reduced photoevaporation rates with that of figure \ref{fig:overview_disc_with_PE} for the nominal values. The reason for this is that the disc in the case of reduced photoevaporative mass loss rates has more time to evolve without gap opening, resulting in a decay of the water content of the inner disc due to accretion onto the host star. Consequently, the effect of the water equilibrium cycle is not as strong as for our nominal simulations, see figure \ref{fig:overview_disc_with_PE}. For the $0.3 \, \text{M}_{\odot}$ star, on the other hand, the inner disc is dominated by a CO$_2$ equilibrium cycle after the gap is fully opened. The C/O ratio in this case decreases very slowly, approaching a value of $0.5$. \\
Simulations of discs around low-mass stars that follow a pure viscous evolution \citep[see][]{mahCloseiceLinesSuperstellar2023} and observations of such systems \citep[see e.g.][]{kanwarMINDSHydrocarbonsDetected2024} indicate high C/O ratios of C/O $>1$ of the inner disc, which is not in agreement with the low C/O ratios we obtain from our simulations using the nominal photoevaporation rates from section \ref{ssec:Nominal_values_for_photoevaporative_mass_loss}. However, already the reduction of the photoevaporative mass loss rates by a factor $3-4$ that is studied in this subsection results in C/O ratios that align much better with the observational values. \cite{kanwarMINDSHydrocarbonsDetected2024} find a C/O ratio $>1$ for the system Sz 28, which has a host star mass of $M_{\star} = 0.12 \, \text{M}_{\odot}$. This result is in very good agreement with the C/O ratio we obtain for the inner disc of the $0.1 \, \text{M}_{\odot}$ star between $2-6 \, \text{Myr}$ of its evolution, see lower right panel of figure \ref{fig:overview_disc_with_factor_3to4_lower_PE}. From this, we conclude that a factor $3-4$ reduced photoevaporation rate can already solve the discrepancy between models and observations of discs around low-mass stars.

\subsubsection{Influence of photoevaporation rate vs. disc viscosity on the C/O ratio}
In figure \ref{fig:C_to_O_ratio_comparison}, we plot exemplarily the C/O ratio for the $0.5 \, \text{M}_{\odot}$ star for different scenarios to study the effects of the different photoevaporation strengths in comparison with the influence of the turbulence parameter $\alpha$/the disc's viscosity. The left column shows the C/O ratio for a disc with a viscous parameter of $\alpha = 10^{-4}$, which is the nominal value, and the right column depicts the same disc but with an $\alpha$ parameter of $\alpha = 10^{-3}$. Nominal values of the photoevaporation rate are used for the top row, whereas the bottom row is simulated with the before-mentioned reduced photoevaporation rates. \\
The comparison shows that higher disc viscosities and higher photoevaporation rates lead to lower C/O ratios in the later evolution stages of the inner disc. For the higher disc viscosities, the low C/O ratio is a result of the evaporation lines being shifted outwards in the disc due to enhanced viscous heating. Consequently, no water equilibrium cycle occurs as the water-ice line is now lying in the gap instead of in the inner disc, see figure \ref{fig:overview_disc_with_PE_larger_alpha} for more details on the photoevaporative disc with a higher viscosity. Additionally, the inner disc is accreted much faster onto the host star as compared to the low-viscosity case. Therefore, the C/O is low after gap opening. For higher photoevaporation rates, the C/O is low due to a different reason. In this case, photoevaporative gaps open fast, leaving no time for carbon-rich vapour to reach the inner disc to enhance the C/O ratio. A reduced photoevaporative mass loss rate, in combination with a viscous parameter of $\alpha = 10^{-4}$, and the subsequent delayed opening of gaps is therefore the only case, where an elevated C/O ratio in the later disc evolution stages is possible, see lower left panel of figure \ref{fig:C_to_O_ratio_comparison}.

\subsection{Disc lifetimes} \label{ssec:Disc_lifetimes}
\begin{figure}
    \centering
    \begin{minipage}{0.5\textwidth}
        \includegraphics[width=\textwidth]{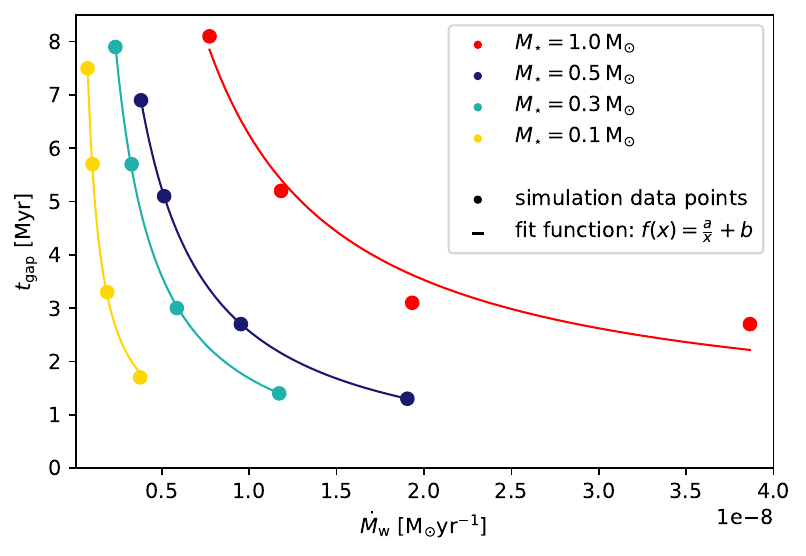}
    \end{minipage}
    \caption{Gap opening time vs. photoevaporation rate for the low-mass stars studied in this paper as well as the solar-mass star from Paper I (colour coded). Plotted are data points from our simulations, both from nominal (rightmost point of each line) and reduced photoevaporative mass loss rates with reduction factors of $2$, $3-4$ and $5$, and a reciprocal fit through those points. The fit parameters for each stellar mass are listed in table \ref{table:fit_parameters_disc_lifetime}.}
    \label{fig:disc_lifetimes}
\end{figure}

\begin{table}
    \centering
	\begin{tabular}{ccc}
		\hline
		$M_{\star}$ $[\text{M}_{\odot}]$	& $a$ $[10^{-2} \, \text{M}_{\odot}]$ & $b$ [Myr]	\\
		\hline \hline
            $1.0$	& $5.44$	& $\phantom{-}0.81$	\\
            $0.5$	& $2.67$	& $-0.10$	\\
            $0.3$	& $1.92$	& $-0.23$	\\
            $0.1$	& $0.55$	& $\phantom{-}0.33$	\\ \hline
	\end{tabular}
    \caption{Fit parameters for different stellar masses for the reciprocal fit function $f(x) = \frac{a}{x} + b$, which fits the simulation data points of the gap opening time as a function of photoevaporation rate, as shown in figure \ref{fig:disc_lifetimes}.}
    \label{table:fit_parameters_disc_lifetime}
\end{table}

For a more sophisticated analysis and comparison of the gap opening times depending on the photoevaporative mass loss rates, we added simulations to our data set with general reduction factors of $2$, $5$ and $10$ compared to the nominal values. In figure \ref{fig:disc_lifetimes}, we show these together with the nominal values from section \ref{ssec:Nominal_values_for_photoevaporative_mass_loss} and the individual reduced mass loss rates from section \ref{ssec:Reduced_photoevaporation_rate} for our stellar sample of low-mass stars studied in this paper as well as the solar-mass star from Paper I. However, only four data points instead of the expected five are visible in figure \ref{fig:disc_lifetimes} for each stellar mass because a reduction of the photoevaporative mass loss rates by a factor of $10$ results in discs that do not show any signs of a photoevaporative gap. Therefore, they are not included in this figure. \\
We define the gap opening time as the time when the gas surface density drops below a specific threshold of $\Sigma_{\text{thresh}} = 10^{-8} \, \text{g/cm}^2$ in the area between $0.1 \, \text{AU}$ and the position of the CO ice line of the respective disc. Figure \ref{fig:disc_lifetimes} depicts a comparison between the different host star masses, which are indicated via colour coding. The gap opening times are plotted as a function of the photoevaporation rate, disc lifetimes can be inferred from the gap opening times. Our simulation results are shown as data points in the figure. Additionally, we plot a reciprocal fit function through the data points for each stellar mass. The corresponding fit parameters can be found in table \ref{table:fit_parameters_disc_lifetime}. \\
In figure \ref{fig:guedel_fig_12}, we have added the nominal photoevaporation rates and the ones with reduction factors of $2$, $3-4$, $5$ and $10$ to the observational data, after converting these rates to X-ray luminosities. The gap opening times for each case are highlighted with colour. \\
It becomes evident that the photoevaporation rate is a critical parameter as the gap opening time varies between around $1-8 \, \text{Myr}$ when changing said photoevaporation rate, depending on the host star mass. The gap opening time then directly influences the chemical evolution of the inner disc, resulting in different C/O ratios, with a later gap opening resulting in higher C/O ratios of the inner disc.

\section{Discussion} \label{sec:Discussion}
\subsection{Dependence on model assumptions} \label{ssec:Dependence_on_model_assumptions}
As we have seen in sections \ref{ssec:Nominal_values_for_photoevaporative_mass_loss} and \ref{ssec:Reduced_photoevaporation_rate}, the results of our simulations strongly depend on the photoevaporative mass loss rate. However, this is not the only parameter that influences the outcome of our simulations. In this section, we qualitatively discuss the dependence of some of our model assumptions. For more details, we refer the interested reader to the discussion in Paper I.

\subsubsection{Photoevaporation rate}
The comparison of our results in section \ref{ssec:Nominal_values_for_photoevaporative_mass_loss}, where we use the nominal values from \cite{picognaDispersalProtoplanetaryDiscs2021b} for the photoevaporative mass loss, to our results in section \ref{ssec:Reduced_photoevaporation_rate}, where the photoevaporative mass loss is reduced by a factor $3-4$, already shows how sensitive to this parameter the inner disc's chemistry is. Note that the reduction of the photoevaporation rate is motivated by a large spread in the X-ray luminosity of stars, as indicated by observations, see figure \ref{fig:guedel_fig_12}. A result of the lower photoevaporation rates is a delay in the gap opening time, leading to higher C/O ratios in the inner disc. This is essential as the nominal photoevaporation rates overestimate the mass loss, resulting in lower C/O ratios which are not in agreement with observations of discs around low-mass stars \citep[see e.g.][]{taboneRichHydrocarbonChemistry2023,kanwarMINDSHydrocarbonsDetected2024}. \\
In addition to the indication by observations, simulations by \cite{sellekPhotoevaporationProtoplanetaryDiscs2024} show that lower photoevaporation rates should be favoured over those by e.g. \cite{picognaDispersalProtoplanetaryDiscs2021b}. Lower photoevaporative mass loss rates in the simulations of \cite{sellekPhotoevaporationProtoplanetaryDiscs2024} result from additional cooling processes which have not been considered in the model of \cite{picognaDispersalProtoplanetaryDiscs2021b}. \\
The sensitivity of the inner disc's chemistry to the photoevaporative mass loss rates becomes even more evident when looking at the simulation results of a disc with a factor $10$ reduced photoevaporation rate, see appendix \ref{asec:Photoevaporative_disc}. We see that a reduction factor of $10$ already reduces the photoevaporation rates so much that we do not observe a significant variation in the disc structure during our studied evolution of $10 \, \text{Myr}$. The discs in these simulations do not show any signs of gap opening, with the results therefore looking almost identical to those of purely viscously evolving discs, compare figure \ref{fig:overview_disc_with_factor_10_lower_PE} to figure \ref{fig:overview_viscous_disc}.

\subsubsection{Photoevaporation of different molecules}
Internal photoevaporation is implemented in our code \texttt{chemcomp} in a way that it acts uniformly on all molecules present in the code. As a result, all species are evaporated equally and leave the disc via photoevaporative winds on the same timescales. However, in reality, it is physically more accurate that lighter molecules are removed more efficiently than molecules with a larger mass, e.g. it is easier for a photoevaporative wind to blow away H$_2$ rather than H$_2$O. This results in different disc leaving timescales for different molecules. However, it is very difficult to assess how this change in timescales would affect the evolution of the inner disc chemistry. To evaluate this, a detailed calculation of the disc leaving timescales for the different molecules present in \texttt{chemcomp} would be needed.

\subsubsection{Refractory carbon grains}
Refractory carbon is not included in the molecule list used for the simulations done in this paper. Its inclusion might however impact the C/O ratio of the inner disc. One possibility to take refractory carbon into account is the existence of a so-called soot line (\textasciitilde$\, 300 \, \text{K}$), where carbon grains then sublimate at much lower temperatures than the actual sublimation temperature of carbon ($2000 \, \text{K}$). This has been proposed by \cite{vanthoffCarbongrainSublimationNew2020} in their work. However, the soot line requires heavily modified carbonaceous material normally occurring in meteoritic parent bodies and does not reflect the state of carbonaceous grains in discs. A second possibility is the destruction of carbon grains via pyrolysis and oxidation, releasing carbon into the gas phase at lower temperatures as well \citep{leeSOLARNEBULAFIRE2010,gailSpatialDistributionCarbon2017,weiEffectCarbonGrain2019}. The third possibility is the reaction of carbon grains with hydrogen to form small molecules like CH$_4$ or C$_2$H$_2$, which would also happen at lower temperatures \citep{nakanoEvaporationInterstellarOrganic2003,liEarthsCarbonDeficit2021,kanwarHydrocarbonChemistryInner2024,raulTrackingChemicalEvolution2025}. \cite{lenzuniDustEvaporationProtostellar1995} and \cite{borderiesDustEvolutionChemisputtering2025} have shown that the latter is the most efficient of these processes. In the first two cases of a soot line and carbon destruction via pyrolysis and oxidation, the freed carbon will also react with hydrogen. The formation of molecules like CH$_4$ or C$_2$H$_2$ might hinder the appearance of an equilibrium cycle for water or CO$_2$. On the other hand, including refractory carbon via one of the above-mentioned processes might lead to an elevation of the C/O ratio of the inner disc as more carbon is then available via CH$_4$ and C$_2$H$_2$.

\subsubsection{Initial disc radius}
\cite{banzattiJWSTRevealsExcess2023} indicate that the water abundance in the inner disc depends on the outer radius of the disc, which is defined by the location of the outermost gap. For this study, they analyse the spectra of four discs observed with the \ac{JWST} \ac{MIRI}, with two of them being compact discs ($10 - 20 \, \text{AU}$) and two of them being large discs ($100 - 150 \, \text{AU}$) with multiple gaps. The two compact discs show a water excess in comparison to the extended discs. \\
On the other hand, different \ac{JWST} observations reveal a large water reservoir in an extended disc with substructures \citep{gasmanMINDSAbundantWater2023}, which may be an indication for gap structures playing a more dominant role in creating water reservoirs than the disc radius \citep[for a discussion of an extended set of sources, see][]{gasmanMINDSInfluenceOuter2025}. As we have already discussed in Paper I for solar-mass stars, our simulations support the latter claim. This is also true for low-mass stars. On the one hand, we find in the simulations done in this paper that the disc radius does not affect the evolution of the gas surface density and the C/O ratio of lower-mass stars, see appendix \ref{asec:Pure_viscous_disc} and compare figure \ref{fig:overview_viscous_disc} with figure \ref{fig:overview_disc_with_smaller_initial_disc_radii}. On the other hand, pebbles in the outer areas of extended discs generally grow more slowly and reach the inner disc later than those in discs with smaller radii. The inner disc of an extended disc is therefore enriched with less water vapour over a longer period in comparison to the inner region of a compact disc. Smaller discs have in turn a higher water content over a shorter period because the pebbles grow faster in the outer disc regions due to a higher density and then have an enhanced drifting speed towards the inner disc. However, in both cases, the inner disc is enriched with water vapour, contradicting the argumentation of the disc radius being the sole indicator of a water-rich disc. Far more important for the chemical evolution of the inner disc is the presence of gaps and their opening timescales as we have shown in sections \ref{ssec:Nominal_values_for_photoevaporative_mass_loss} and \ref{ssec:Reduced_photoevaporation_rate}.

\subsection{Implications of our results} \label{ssec:Implications_of_our_results}
\ac{JWST} observations of the \ac{MINDS} collaboration of the system Sz 28 ($M_{\star} = 0.12 \, \text{M}_{\odot}$) suggest a high gaseous C/O ratio of C/O $>1$ for the inner disc, see \cite{kanwarMINDSHydrocarbonsDetected2024}. Constraining the C/O ratio in the inner disc of PDS 70 ($M_{\star} = 0.76 \, \text{M}_{\odot}$) also predicts a super-stellar C/O ratio, see \cite{portillareveloClosingGapTheory2023}. Other papers on low-mass stars, see \cite{taboneRichHydrocarbonChemistry2023} or \cite{arabhaviAbundantHydrocarbonsDisk2024}, show results going in the same direction although these do not include detailed thermochemical calculations. \ac{ALMA} observations on the other hand mostly give estimates on the C/O ratio of the outer disc ($> 10 \, \text{AU}$) \citep[see e.g.][]{bosmanMoleculesALMAPlanetforming2021,legalMoleculesALMAPlanetforming2021}. For most discs, there is no \ac{ALMA} data available probing the gas in the inner disc ($< 10 \, \text{AU}$). \\
These observational results imply that our nominal values for the photoevaporative mass loss are indeed overestimating the actual mass loss, leading to C/O ratios in the inner disc that are too low to match observations, see section \ref{ssec:Nominal_values_for_photoevaporative_mass_loss}. The reduced rates discussed in section \ref{ssec:Reduced_photoevaporation_rate} on the other hand fit the observational values much better. Especially, the supersolar C/O ratio of the $0.1 \, \text{M}_{\odot}$ star between $2-6 \, \text{Myr}$ matches very well the observed values for the system Sz 28 \citep{kanwarMINDSHydrocarbonsDetected2024}.

\section{Summary and conclusions} \label{sec:Summary_and_conclusions}
In the second paper of our series, we perform 1D semi-analytical simulations of protoplanetary discs around low-mass stars, with masses ranging from $0.1 \, \text{M}_{\odot}$ to $0.5 \, \text{M}_{\odot}$. Our model includes pebble drift and evaporation. Additionally, internal photoevaporation due to X-rays is active. Combining both these mechanisms opens up new perspectives for understanding the composition and evolution of inner discs. We compare our results, where we use nominal values for the photoevaporative mass loss from \cite{picognaDispersalProtoplanetaryDiscs2021b}, to discs with less effective photoevaporation. \\
Our results clearly indicate that internal photoevaporation strongly affects the evolution and chemical composition of protoplanetary discs. Due to the consequential gap opening, gas diffusing into such gaps is carried away by photoevaporative winds and pebbles from the outer regions are blocked in their inward motion. \\
For the nominal photoevaporation rates, this results in early gap opening and a low C/O ratio in the inner disc, where the latter contrasts observations of discs around low-mass stars. Our model can naturally explain the observed high C/O ratios of C/O $>1$ when using factor $3-4$ reduced photoevaporative mass loss rates, which lead to delayed gap opening times. Using the nominal mass loss rates on the other hand results in the removal of the discs after $1-2 \, \text{Myr}$, which is in agreement with the observations of stars that have already lost their discs at such a young age \citep{pfalznerMostPlanetsMight2022}. \\
This dichotomy leads to the following implications for low-mass stars: Our model predicts that inner discs of young low-mass stars ($< 2 \, \text{Myr}$) should be oxygen-rich and carbon-poor, while older discs ($> 2 \, \text{Myr}$) should be carbon-rich. The latter need to have lower photoevaporation rates, where such low rates either originate from the large spread of observed X-ray luminosities or the photoevaporation model used in this study \citep{picognaDispersalProtoplanetaryDiscs2021b}. This model likely overestimates the photoevaporative mass loss at a given X-ray luminosity, leading to a mismatch between the calculated and the observed C/O ratios in discs around low-mass stars. A reduction of the photoevaporation rate by a factor of $3-4$ brings the elemental abundances from our simulations into better agreement with observations.

\section*{Acknowledgements}
We thank Giulia Perotti for helpful discussions and Manuel Güdel for providing the data to create figure \ref{fig:guedel_fig_12}. \\
Th.H. acknowledges the support of the ERC Origins grant number 832428. J.L.L. is a fellow of the International Max Planck Research School for Astronomy and Cosmic Physics at the University of Heidelberg (IMPRS-HD).

\bibliographystyle{aa}
\bibliography{Paper_2}

\begin{appendix}
\section{Additional material: Pure viscous disc} \label{asec:Pure_viscous_disc}
\begin{figure*}
    \centering
    \includegraphics[width=\textwidth]{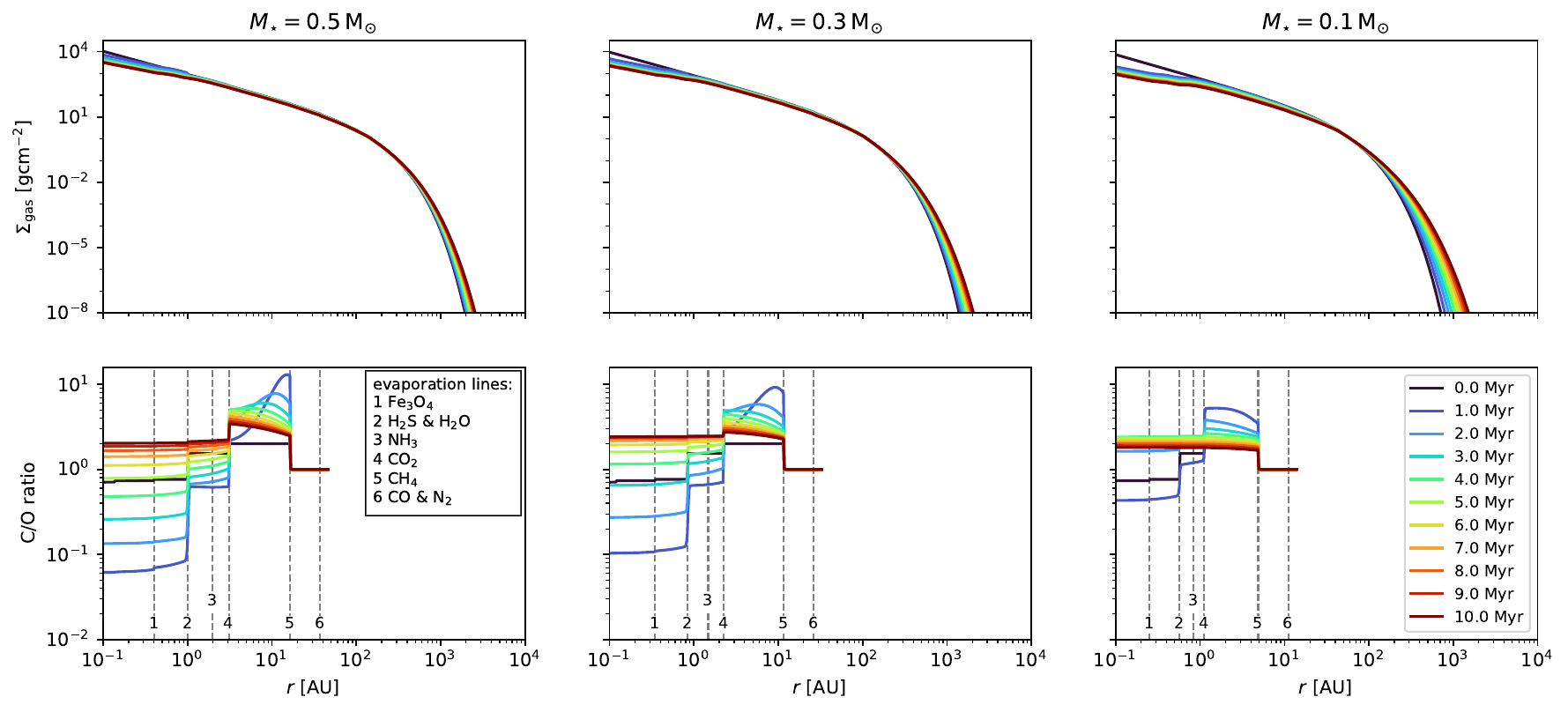}
    \caption{Disc evolution for a viscous disc without internal photoevaporation. The host star masses vary from $0.5 \, \text{M}_{\odot}$ (on the left) to $0.1 \, \text{M}_{\odot}$ (on the right). \textbf{Top:} Gas surface density as a function of disc radius and time. \textbf{Bottom:} Gaseous C/O ratio as a function of disc radius and time. Colour coding, plotting and simulation parameters as in figure \ref{fig:overview_disc_with_PE}.}
    \label{fig:overview_viscous_disc}
\end{figure*}

\begin{figure*}
    \centering
    \includegraphics[width=\textwidth]{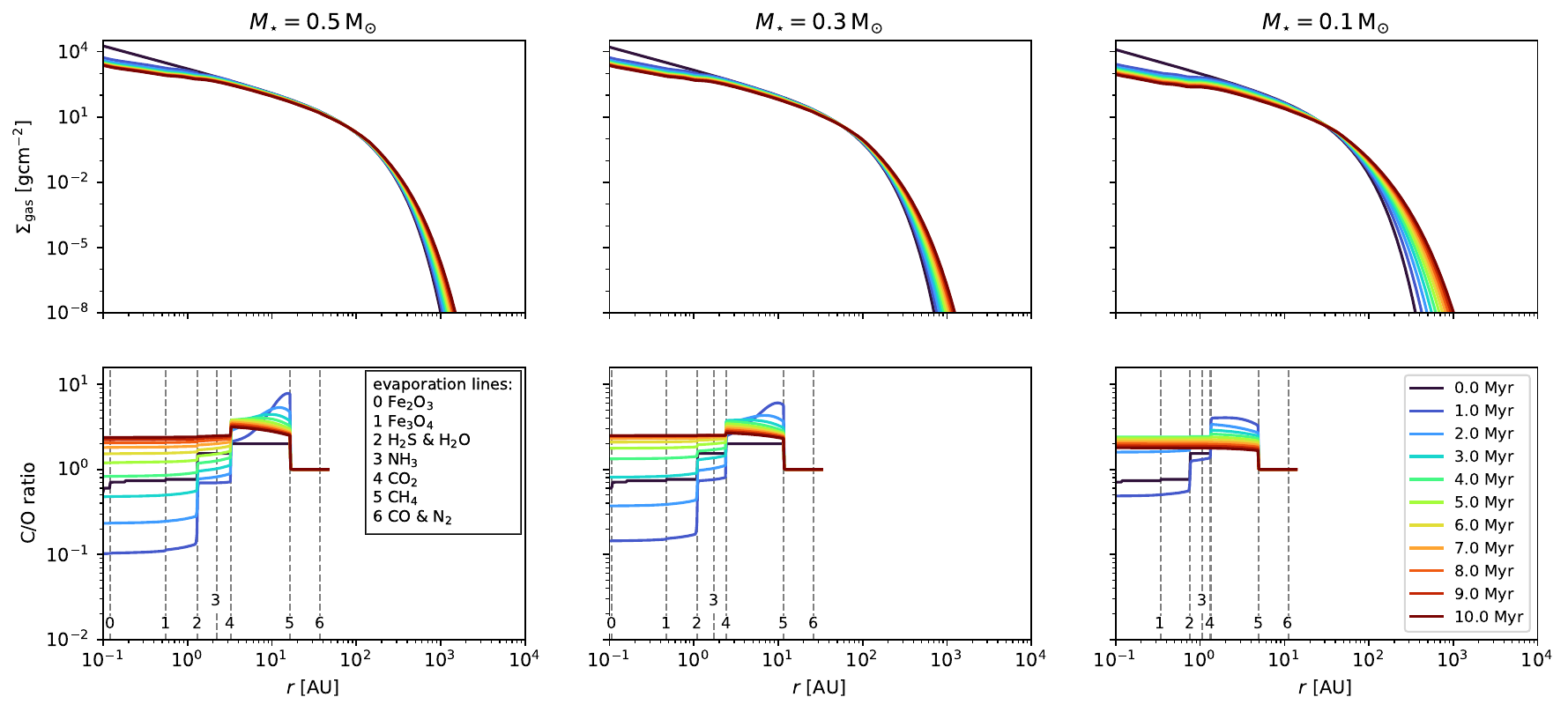}
    \caption{Disc evolution for a viscous disc without internal photoevaporation. The initial disc radii are reduced by a factor of $2$ compared to the nominal values given in table \ref{table:simulation_parameters}. Their values are now $[47.5,32.5,15] \, \text{AU}$ for our host star masses of $0.5 \, \text{M}_{\odot}$ (left), $0.3 \, \text{M}_{\odot}$ (middle) and $0.1 \, \text{M}_{\odot}$ (right). \textbf{Top:} Gas surface density as a function of disc radius and time. \textbf{Bottom:} Gaseous C/O ratio as a function of disc radius and time. Colour coding, plotting and remaining simulation parameters as in figure \ref{fig:overview_disc_with_PE}.}
    \label{fig:overview_disc_with_smaller_initial_disc_radii}
\end{figure*}

\subsection{Nominal case} \label{assec:Nominal_case}
The results for the pure viscous disc are obtained using the standard simulation parameters as described in section \ref{ssec:Initial_conditions} and shown in table \ref{table:simulation_parameters}. In this case, internal photoevaporation is switched off to have a comparison disc which is dominated solely by viscous evolution. \\
In figure \ref{fig:overview_viscous_disc}, we show the gas surface density in the top row and the C/O ratio in the bottom row, both as a function of disc radius and time. The data is generated and plotted as in the case with photoevaporation, see section \ref{ssec:Nominal_values_for_photoevaporative_mass_loss} for a detailed description. Variation in stellar mass is indicated at the top of each column, with the masses decreasing from $0.5 \, \text{M}_{\odot}$ on the left to $0.1 \, \text{M}_{\odot}$ on the right. \\
Our results for the gas surface density in the top row of figure \ref{fig:overview_viscous_disc} do not show a significant difference between the different host star masses. We see a decrease in the gas surface density in the inner disc for all three cases, resulting from the viscous accretion of disc material onto the central star. Additionally, the outer disc is dominated by viscous spreading. These two effects increase in their intensity with decreasing host star mass and happen on relatively long timescales as the viscous parameter $\alpha$ is small, $\alpha = 10^{-4}$. The results obtained for the gas surface density align with those from Paper I, where we studied solar-mass stars. \\
The C/O ratio, depicted in the bottom row of figure \ref{fig:overview_viscous_disc}, also shows similar behaviour to that of the inner discs around solar-mass stars, except for the $0.1 \, \text{M}_{\odot}$ star. For all three host star masses studied here, the C/O ratio drops during the first one million years, followed by an increase. In the case of the $0.5 \, \text{M}_{\odot}$ and $0.3 \, \text{M}_{\odot}$ stars, this rise in the C/O ratio continues until the end of the time evolution. For the $0.1 \, \text{M}_{\odot}$ star, however, this rise is followed by a very slow decrease, leading to a more or less constant C/O ratio over time. \\
As in the case of a photoevaporative disc, see section \ref{ssec:Nominal_values_for_photoevaporative_mass_loss}, the behaviour of the C/O ratio is a direct consequence of pebble drift and evaporation. Due to the water-ice line being closer to the star and the fact that pebbles drift much faster through the disc than gas, the inner disc is first enriched with water vapour. This results in high amounts of oxygen being present in the beginning, leading to a low C/O ratio. With carbon-rich vapour, which forms in the outer disc, arriving later to the inner regions, this trend is slowly reversed. As a result, the C/O ratio increases, reaching super-solar values after a few million years. However, in the case of the $0.1 \, \text{M}_{\odot}$ star, the C/O ratio decreases again after around $4 \, \text{Myr}$ while remaining supersolar. This is due to the viscous evolution being faster for smaller discs. The ice lines are much closer to the host star for the $0.1 \, \text{M}_{\odot}$ star than for the $0.5 \, \text{M}_{\odot}$ star. The former is therefore much faster dominated by CH$_4$ and CO, leading to a C/O ratio of about $2$, as CH$_4$ and CO exist in the same fraction in our simulations. On the other hand, the $0.5 \, \text{M}_{\odot}$ star is at the same time still dominated by solely CH$_4$, leading to a higher C/O ratio.

\subsection{Smaller initial disc radius} \label{assec:Smaller_initial_disc_radius}
This section compares our nominal viscous discs to discs with a factor $2$ smaller initial disc radius. This decreases $R_{\text{disc}}$ from $[95,65,30] \, \text{AU}$ to $[47.5,32.5,15] \, \text{AU}$ for our host star masses of $[0.5,0.3,0.1] \, \text{M}_{\odot}$. The rest of the simulation parameters are kept as before, see section \ref{ssec:Initial_conditions} and table \ref{table:simulation_parameters} for more details. \\
Figure \ref{fig:overview_disc_with_smaller_initial_disc_radii} shows the gas surface in the top row and the gaseous C/O ratio in the bottom row, both as a function of disc radius and time. The columns correspond to different host star masses, varying from $0.5 \, \text{M}_{\odot}$ on the left to $0.1 \, \text{M}_{\odot}$ on the right. \\
The general evolution of the gas surface density in discs with smaller initial disc radii is very similar to that of discs with larger initial radii, compare the top rows of figure \ref{fig:overview_viscous_disc} and figure \ref{fig:overview_disc_with_smaller_initial_disc_radii}. However, smaller initial radii lead to faster evolution, which can be seen in the accretion of inner disc material onto the central star and in the viscous spreading in the outer disc. \\
The behaviour of the C/O ratio in a viscous disc with a smaller initial radius, as shown in the bottom row of figure \ref{fig:overview_disc_with_smaller_initial_disc_radii}, is also very similar to that of a disc with a larger initial radius, see bottom row of figure \ref{fig:overview_viscous_disc}.

\section{Additional material: Viscous disc with a planet} \label{asec:Viscous_disc_with_a_planet}
\begin{table}
    \centering
	\begin{tabular}{ccc}
		\hline
		$M_{\star}$ $[\text{M}_{\odot}]$   & $M_{\text{planet}}$ $[\text{M}_{\oplus}]$  & $t_{\text{pim}}$ $[\text{Myr}]$   \\
            \hline\hline
            $0.5$   & $399$             & $0.12$    \\
            $0.3$   & $163$             & $0.14$    \\
            $0.1$   & $\phantom{0}11$   & $0.27$    \\  \hline
	\end{tabular}
    \caption{Planetary masses and timescales for reaching \ac{pim}.}
    \label{table:planet_parameters}
\end{table}

\begin{figure*}
    \centering
    \includegraphics[width=\textwidth]{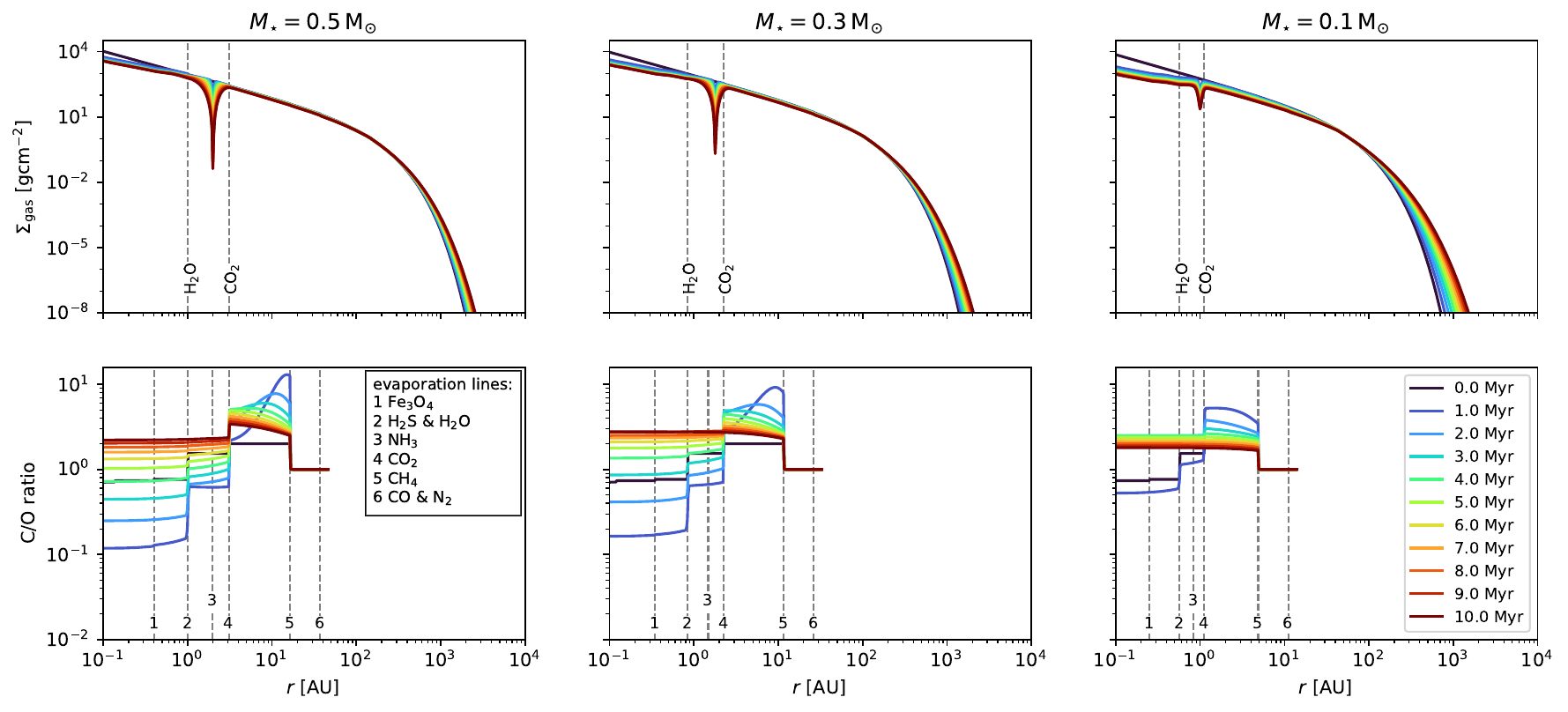}
    \caption{Disc evolution for a viscous disc without internal photoevaporation, where a planet seed is put at $[2.0,1.8,1.0] \, \text{AU}$ at $0.05 \, \text{Myr}$, using an envelope opacity of $\kappa_{\text{env}} = 0.5 \, \text{cm}^2\text{/g}$. The planet has a final mass of about $[400,165,10] \, \text{M}_{\text{Earth}}$ and reaches pebble isolation mass at $[0.12,0.14,0.27] \, \text{Myr}$ for our host star masses of $[0.5,0.3,0.1] \, \text{M}_{\odot}$. The stellar mass decreases from the left to the right. \textbf{Top:} Gas surface density as a function of disc radius and time. \textbf{Bottom:} Gaseous C/O ratio as a function of disc radius and time. Colour coding, plotting and simulation parameters as in figure \ref{fig:overview_disc_with_PE}.}
    \label{fig:overview_disc_with_planet}
\end{figure*}

In addition to the comparison to a viscous disc, it is important to compare the photoevaporative disc to a viscous disc hosting a giant planet. Especially since in both of the latter cases, the disc is dominated by a gap. In figure \ref{fig:overview_disc_with_planet}, we therefore plot the gas surface density in the top row and the C/O ratio in the bottom row for a viscous disc with a growing planet, again as a function of both disc radius and time. We vary the central star's mass, as indicated at the top of each column. \\
The results in this section are obtained using the standard simulation parameters as described in section \ref{ssec:Initial_conditions} and shown in table \ref{table:simulation_parameters}. Additionally, we use an envelope opacity of $\kappa_{\text{env}} = 0.5 \, \text{cm}^2\text{/g}$. The value of the envelope opacity determines the duration of the envelope contraction phase of the planet, with high envelope opacities allowing for slower gas accretion compared to low values because high envelope opacities decrease the cooling rate of the envelope. Planets in our model then grow via pebble and gas accretion, migration is turned off. \\
For each case considered here, a planetary seed is placed in the disc between the H$_2$O and the CO$_2$ evaporation line at $0.05 \, \text{Myr}$. This corresponds to $[2.0,1.8,1.0] \, \text{AU}$ for our host star masses of $[0.5,0.3,0.1] \, \text{M}_{\odot}$. The planetary seeds do not migrate but grow over time due to pebble and gas accretion, where pebble accretion stops at $[0.12,0.14,0.27] \, \text{Myr}$, respectively, when the planets reach pebble isolation mass. At this time, the pebbles are blocked from reaching the inner disc and their inward flux is stopped. However, gas from the outer disc is still able to diffuse through the gap, \citep[see e.g.][]{paardekooperDustFlowGas2006,lambrechtsSeparatingGasgiantIcegiant2014,ataieeHowMuchDoes2018,bitschPebbleisolationMassScaling2018,weberCharacterizingVariableDust2018}, moving towards the inner disc regions. It is only slightly hindered by the growing planet. This distinguishes gaps opened by planets from those caused by internal photoevaporation. In both cases, the pebble flux to the inner disc is stopped once the gap is fully opened/the planet reaches pebble isolation mass. However, in the case of a planetary gap, gas can still move inwards, whereas photoevaporative winds carry away the gas completely, resulting in a cut-off of the inner disc from the outer disc's gas supply. \\
The planets in our simulations reach final masses of about $[400,165,10] \, \text{M}_{\text{Earth}}$, for a list of their exact parameters see table \ref{table:planet_parameters}. Their position is seen as a dip in the gas surface density, see upper row of figure \ref{fig:overview_disc_with_planet}. With time, the dip deepens as the planetary mass grows. \\
The C/O ratio evolves analogously to that of a pure viscous disc without a planet, compare the bottom row of figure \ref{fig:overview_disc_with_planet} to the bottom row of figure \ref{fig:overview_viscous_disc}. The only difference that is occurring when a planet is present is a generally higher C/O ratio in the inner disc throughout the full time evolution. This is because the planets in all scenarios studied here reach pebble isolation mass relatively early, after $[0.12,0.14,0.27] \, \text{Myr}$ for our host star masses of $[0.5,0.3,0.1] \, \text{M}_{\odot}$. After reaching pebble isolation mass, all pebbles from the outer disc are blocked. This is especially true for the water-rich pebbles since the planets are located between the water evaporation front and the ice lines of carbon-bearing molecules. As a result, the water-rich pebbles are trapped in the outer disc with no chance of evaporating their water-ice. Carbon-rich pebbles on the other hand are still able to evaporate their volatile content. Therefore, carbon-rich gas is created outside of the planet's position, can pass the planetary gap and move to the inner disc. Subsequently, water-rich pebbles have much less time than in the purely viscous case to enrich the inner disc with water vapour before gap-opening. Consequentially, the carbon-rich gas, which arrives later, needs to balance less oxygen, ultimately leading to a higher C/O ratio. In the planetary disc, it stays continuously higher than in the pure viscous disc because only water-rich pebbles are blocked but carbon-rich gas can still pass to the inner disc. The result of this process is a supersolar C/O ratio of the inner disc, which is even higher than in the pure viscous disc for the $0.5 \, \text{M}_{\odot}$ and $0.3 \, \text{M}_{\odot}$ stars. However, the $0.1 \, \text{M}_{\odot}$ star shows the same C/O ratio regardless if a planet is forming in the disc or not. The reason is that the C/O ratio in this case towards the end of the time evolution at $10 \, \text{Myr}$ is entirely determined by the CH$_4$ and CO gas moving inwards from the outer disc regions, see the discussion in section \ref{assec:Nominal_case} for more details. \\
Finally, note that a variation in the envelope opacity $\kappa_{\text{env}}$ only changes the final mass of the planets, but not the C/O ratio in the disc. Once a planet reaches pebble isolation mass, the pebbles are blocked by the planetary gap. However, gas can still pass through the gap and move into the inner disc. Depending on the mass of the planet, more or less gas is hindered in its flow, but this does not change the overall C/O ratio as both carbon and oxygen are hindered equally from moving inwards.

\section{Additional material: Photoevaporative disc} \label{asec:Photoevaporative_disc}
\begin{figure*}
    \centering
    \includegraphics[width=\textwidth]{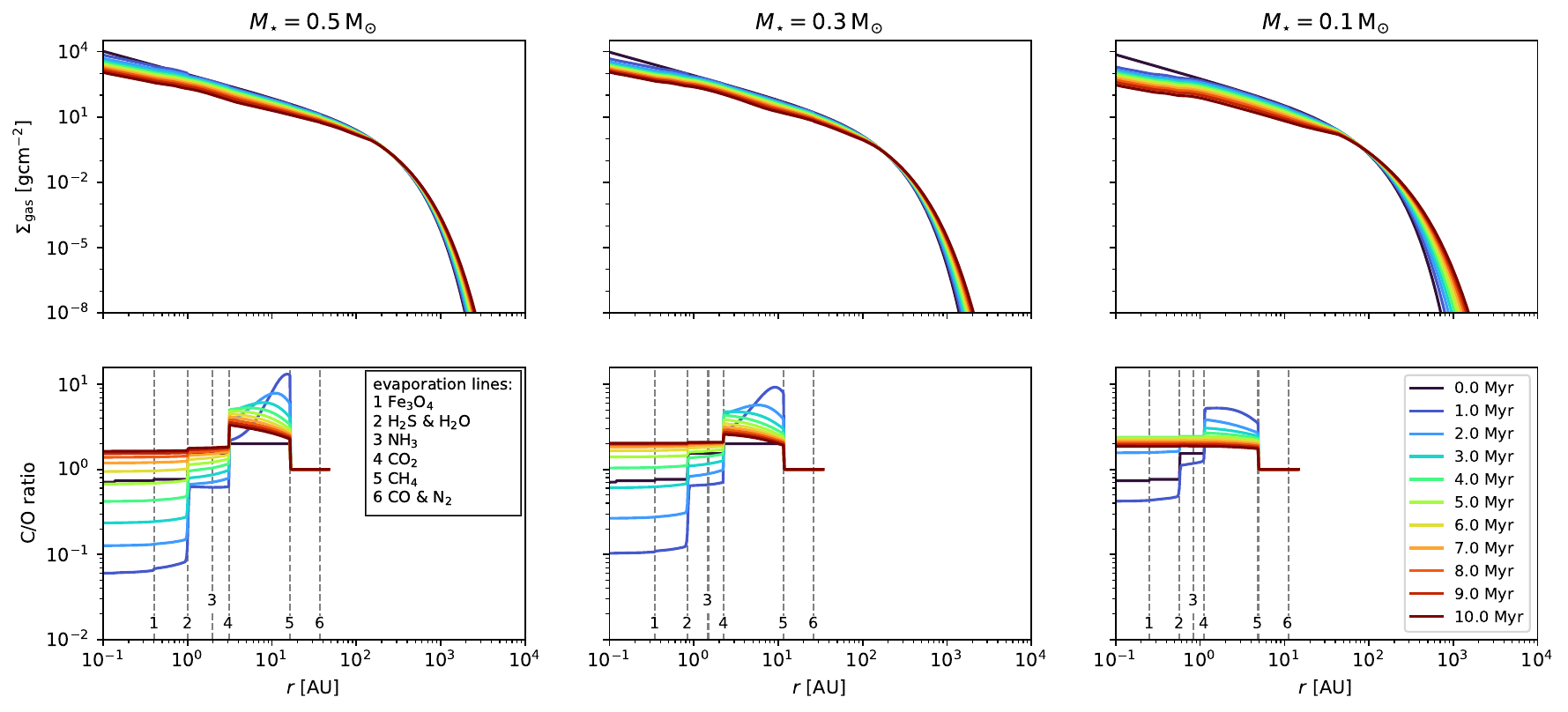}
    \caption{Disc evolution for a viscous disc with internal photoevaporation due to X-rays, but with a factor $10$ reduced mass loss rate. The host star masses vary from $0.5 \, \text{M}_{\odot}$ (on the left) to $0.1 \, \text{M}_{\odot}$ (on the right). \textbf{Top:} Gas surface density as a function of disc radius and time. \textbf{Bottom:} Gaseous C/O ratio as a function of disc radius and time. Colour coding, plotting and simulation parameters as in figure \ref{fig:overview_disc_with_PE}.}
    \label{fig:overview_disc_with_factor_10_lower_PE}
\end{figure*}

\begin{figure*}
    \centering
    \includegraphics[width=\textwidth]{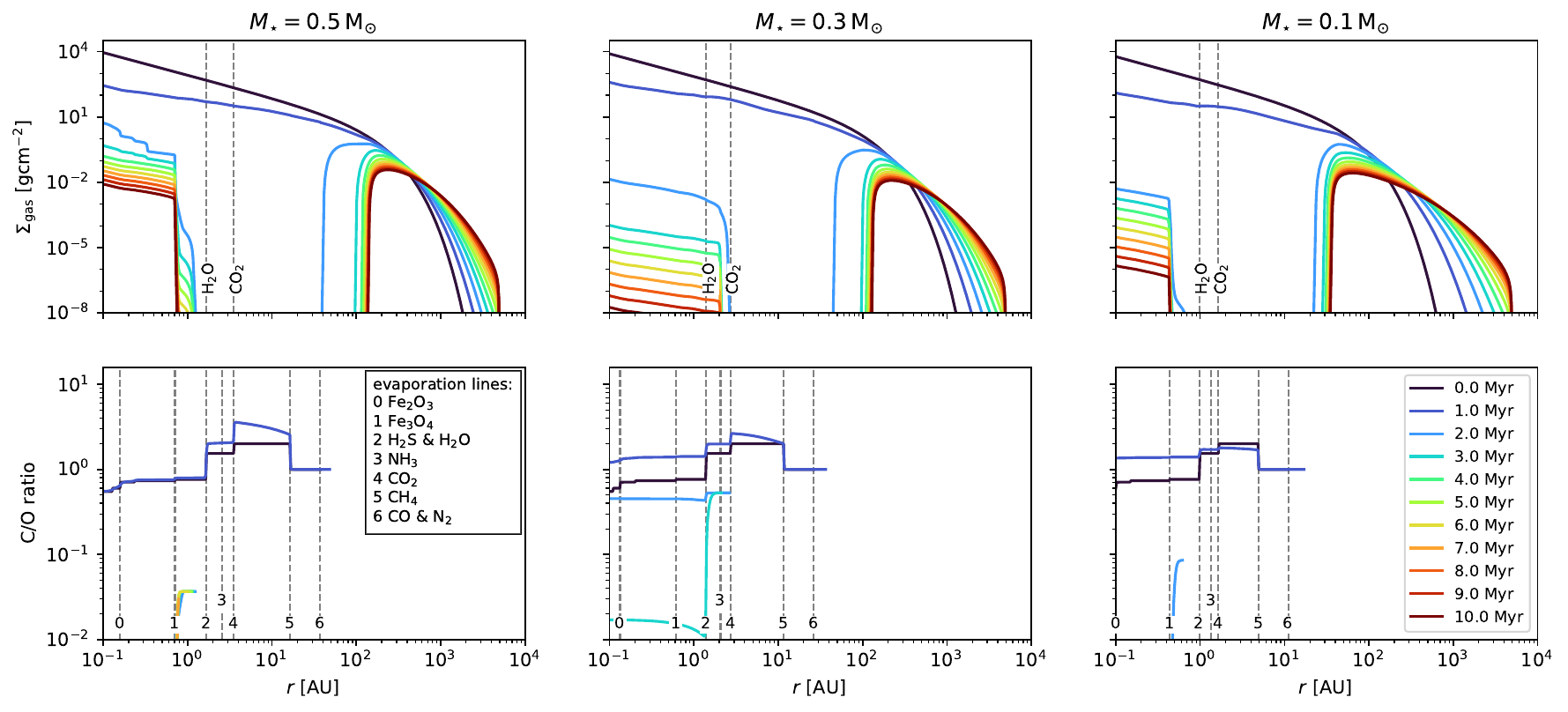}
    \caption{Disc evolution for a viscous disc with internal photoevaporation due to X-rays, using the nominal photoevaporative mass loss rates from table \ref{table:fit_parameters} and an $\alpha$ parameter of $\alpha = 10^{-3}$, corresponding to a factor of $10$ increase compared to our nominal viscosity. The host star masses vary from $0.5 \, \text{M}_{\odot}$ (on the left) to $0.1 \, \text{M}_{\odot}$ (on the right). \textbf{Top:} Gas surface density as a function of disc radius and time. \textbf{Bottom:} Gaseous C/O ratio as a function of disc radius and time. Colour coding, plotting and remaining simulation parameters as in figure \ref{fig:overview_disc_with_PE}.}
    \label{fig:overview_disc_with_PE_larger_alpha}
\end{figure*}

\subsection{Factor 10 reduced mass loss rates}
Here, we show the results for a viscous disc with active internal photoevaporation. The photoevaporative mass loss rate is reduced globally by a factor of $10$ for all host star masses considered here. Our results are obtained using the standard simulation parameters as described in section \ref{ssec:Initial_conditions} and shown in table \ref{table:simulation_parameters}. \\
Figure \ref{fig:overview_disc_with_factor_10_lower_PE} shows the gas surface density in the top row and the C/O ratio in the bottom row, both as a function of disc radius and time. The mass of the central star is varied across the different columns, indicated at the top of each of them, with the masses decreasing from $0.5 \, \text{M}_{\odot}$ on the left to $0.1 \, \text{M}_{\odot}$ on the right. \\
The results for both gas surface density and C/O ratio are consistent with those from a purely viscously evolving disc up to $10 \, \text{Myr}$, see appendix \ref{asec:Pure_viscous_disc}. This consequently follows from the fact that such a low photoevaporation rate does not open any gaps in the disc, leaving the disc no other choice but to follow a pure viscous evolution. However, we expect a gap to open at later times because we already see a small reduction in the gas surface density up to $100 \, \text{AU}$ in the case of the factor $10$ reduced photoevaporation rates when comparing the top rows of figures \ref{fig:overview_viscous_disc} and \ref{fig:overview_disc_with_factor_10_lower_PE}. \\
We therefore do not discuss the results for the disc with a factor $10$ reduced photoevaporative mass loss rate here, the interested reader is referred to appendix \ref{asec:Pure_viscous_disc} for a detailed description. The missing gaps are also the reason why the discs studied here do not appear in the plot in section \ref{ssec:Disc_lifetimes}, where we discuss disc lifetimes, defined via gap opening times, as a function of photoevaporative mass loss rates, see figure \ref{fig:disc_lifetimes}.

\subsection{Larger viscous parameter $\alpha$} \label{assec:Larger_viscous_parameter_alpha}
This section compares our nominal photoevaporative discs to discs with a factor $10$ larger turbulence parameter $\alpha$, we increase $\alpha$ from the nominal value of $10^{-4}$ to $10^{-3}$. The rest of the simulation parameters are the same as before, see section \ref{ssec:Initial_conditions} and table \ref{table:simulation_parameters} for more details. \\
Figure \ref{fig:overview_disc_with_PE_larger_alpha} shows the gas surface density in the top row and the gaseous C/O ratio in the bottom row, both as a function of disc radius and time. The columns correspond to different host star masses, varying from $0.5 \, \text{M}_{\odot}$ on the left to $0.1 \, \text{M}_{\odot}$ on the right. \\
The general evolution of the gas surface density in a disc with a larger turbulence parameter $\alpha$ and a therefore higher viscosity is similar to that of a disc with a lower $\alpha$, compare the top row of figure \ref{fig:overview_disc_with_PE_larger_alpha} to that of figure \ref{fig:overview_disc_with_PE}. However, larger viscosities lead to smaller particles and a faster evolution, both for the accretion of inner disc material onto the central star and the viscous spreading in the outer disc. \\
The C/O ratio in a photoevaporative disc with $\alpha = 10^{-3}$, as shown in the bottom row of figure \ref{fig:overview_disc_with_PE_larger_alpha}, is solar ($M_{\star} = 0.5 \, \text{M}_{\odot}$) or even supersolar ($M_{\star} = 0.3/0.1 \, \text{M}_{\odot}$) before gap-opening but quickly and drastically decreases after the photoevaporative gap has opened. The amount of oxygen in comparison to carbon is so large that after $1 \, \text{Myr}$, for the $0.5/0.1 \, \text{M}_{\odot}$ stars, or $3 \, \text{Myr}$, for the $0.3 \, \text{M}_{\odot}$ star, no C/O ratio can be calculated any more in the parameter range shown here. The high C/O ratios in the first one million years are a result of the much faster evolution in the case of higher viscosities. At this point, the water-rich phase in the inner disc is already over and carbon-rich vapour has already enriched it enough for the C/O ratio to reach such high values. Additionally, high viscosities result in smaller pebbles that then drift inwards slower than larger pebbles. Therefore, the water enrichment of the inner disc in the case of $\alpha = 10^{-3}$ is not as strong as in the case of $\alpha = 10^{-4}$, resulting in less water vapour that needs to be balanced by carbon-rich gas.

\end{appendix}

\end{document}